\documentclass[%
 aip,
 amsmath,amssymb,
 reprint,%
]{revtex4-1}

\usepackage{graphicx}
\usepackage{dcolumn}
\usepackage{bm}

\usepackage[utf8]{inputenc}
\usepackage[T1]{fontenc}
\usepackage{mathptmx}
\usepackage{etoolbox}

\usepackage{color}
\usepackage{float}
\usepackage[hidelinks=true]{hyperref}
\hypersetup{
  colorlinks   = true, 
  urlcolor     = blue, 
  linkcolor    = blue, 
  citecolor   = red 
}

\makeatletter
\def\@email#1#2{%
 \endgroup
 \patchcmd{\titleblock@produce}
  {\frontmatter@RRAPformat}
  {\frontmatter@RRAPformat{\produce@RRAP{*#1\href{mailto:#2}{#2}}}\frontmatter@RRAPformat}
  {}{}
}%
\makeatother
\begin{document}

\preprint{AIP/123-QED}

\title{Accuracy and limitations of the bond polarizability model in  modeling of Raman scattering from molecular dynamics simulations } 
\author{Atanu Paul}
 \affiliation{Department of Chemistry, Bar-Ilan University, Ramat Gan 5290002, Israel}
\author{Maya Rubenstein}%
\affiliation{Department of Chemistry, Bar-Ilan University, Ramat Gan 5290002, Israel}
x\author{Anthony Ruffino}
\affiliation{Department of Physics, Drexel University, Philadelphia, Pennsylvania 19104, USA}
\author{Stefan Masiuk}
\affiliation{Department of Mechanical Engineering and Mechanics, Drexel University, Philadelphia, Pennsylvania 19104, USA}
\author{Jonathan Spanier}
\affiliation{Department of Physics, Drexel University, Philadelphia, Pennsylvania 19104, USA}
\affiliation{Department of Mechanical Engineering and Mechanics, Drexel University, Philadelphia, Pennsylvania 19104, USA}

\author{Ilya Grinberg}
\email[]{ilya.grinberg@biu.ac.il}
\affiliation{%
Department of Chemistry, Bar-Ilan University, Ramat Gan 5290002, Israel
}%


\begin{abstract}
Calculation of Raman scattering from molecular dynamics (MD) simulations requires accurate modeling of the evolution of the electronic polarizability of the system along its MD trajectory. For large systems, this necessitates the use of atomistic models to represent  the dependence of electronic polarizability on atomic coordinates. The bond polarizability model (BPM) is the simplest such model and has been used for modeling the Raman  spectra of molecular systems but has not been applied to solid-state systems.  Here, we systematically investigate the accuracy and limitations of the BPM parameterized from density functional theory (DFT) results for a series of simple molecules such as CO$_{2}$, SO$_{2}$, H$_{2}$S, H$_{2}$O, NH$_{3}$, and CH$_{4}$, the more complex CH$_{2}$O, CH$_{3}$OH and CH$_{3}$CH$_{2}$OH and thiophene molecules and the BaTiO$_{3}$ and CsPbBr$_{3}$ perovskite solids. We find that BPM can reliably reproduce the overall features of the  Raman spectra such as shifts of peak positions. However, with the exception of highly symmetric systems, the assumption of non-interacting bonds limits the quantitative accuracy of the BPM; this assumption also leads to qualitatively inaccurate polarizability evolution and Raman spectra  for systems where large deviations from the ground state structure are present. 
\end{abstract}

\maketitle

\section{INTRODUCTION}
Spectroscopy is as a powerful tool that has been  widely applied to various inorganic and organic systems to obtain the characteristic vibrations and other related properties~\cite{chemosensors9090262,long}. 
The application of  molecular dynamics (MD) simulations  for the interpretation of infrared (IR) spectra has provided a powerful method for detailed understanding of experimental IR spectroscopy results, significantly enhancing the value provided by IR spectra~\cite{C7SC02267K,IR_MD}.  By contrast,  obtaining a similar level of microscopic insight from Raman scattering with the help of MD simulations has not been possible due to the difficulties in the evaluation of the electronic polarizability trajectory.

The Raman spectrum of a system owing to its lattice dynamics can be calculated from  the changes in the electronic polarizability of the system due to its atomic (or ionic) vibrations. Therefore, to model the Raman spectra using MD simulations, electronic polarizability  along the trajectory of the system dynamics must be calculated, and then the Raman spectrum  can be obtained from the auto-correlation function of the time-dependent electronic polarizability trajectory~\cite{Raman_intensity,PhysRevLett.88.176401,hutter}. Electronic polarizability is an observable related to the electronic response of the system to the perturbation by an applied electric field, and ideally should be obtained by quantum-mechanical (QM) calculations~\cite{acs.jctc.0c00755,PhysRevLett.90.036401}.  However, such QM  calculation of electronic polarizability at each time step has a very high computational cost. While atomistic models using neural networks for estimation of electronic polarizability trajectories have been developed, due to their complexity they still have a high computational cost, albeit much smaller one than that of direct QM calculations, and require large amount of QM data for model training~\cite{LUSSIER2020115796,car,berger2024raman,Raimbault_2019}.  Therefore, modeling the electronic polarizability dynamics for large systems is still challenging. 

Among the atomistic approaches for electronic polarizability evaluation, the bond polarizability model (BPM)~\cite{BPM1,cardonalight,1996273},  is the simplest and the most computationally efficient.  This  simple model is based on the assumption that the total polarizability of the system is the sum of the polarizability contributions of the individual  bonds. The model also assumes that the polarizability contribution of each bond is a function of the  length of that bond only, i.e. independent bond approximation. Despite these simplistic assumptions, BPM has been applied  and found to be accurate for obtaining Raman spectra in various systems~\cite{C7NR05839J,PhysRevB.71.241402,Luo2015,PhysRevB.63.094305}.  However, it was also found to be inaccurate for some molecular and solid state systems~\cite{ghosez,berger2023polarizability}.  Therefore, it is important to analyze the BPM as the simplest possible electronic polarizability  model to systematically evaluate the limits of its applicability to the study of molecular and solid-state systems and to provide a baseline reference for the development of more sophisticated electronic polarizablity models.

In this paper,  we first applied the BPM to simple molecules such as the linear molecule CO$_{2}$, and several two-element non-linear molecules SO$_{2}$, H$_{2}$S, H$_{2}$O, NH$_{3}$ and  CH$_{4}$ with different bond angles; these molecules are used as model systems to understand the effect of bond angles on the BPM accuracy. We then examined the more complicated multielement CH$_{2}$O and CH$_{3}$OH and CH$_{3}$CH$_{2}$OH and thiophene molecules to understand the impact of various degrees of freedom of molecular structures on larger molecules in the application of BPM. Finally, we applied the BPM to perovskites as  model solid state systems due to their  broad range of  functional properties and rich physics. In particular, we focused on the classic oxide ferroelectric (FE)  BaTiO$_{3}$ (BTO) and halide perovskite CsPbBr$_{3}$ (CPB). For all of these systems, we compare the BPM results for the polarizability and Raman spectra  with those obtained by  direct  density functional theory (DFT) calculations for small systems. Our results show that despite its simplicity, BPM can be useful and accurate for capturing the overall features of the Raman spectra; however, BPM  fails to reproduce fine polarizability changes and some features of Raman spectra when encountering strong asymmetric vibrations and strong anharmonicity. Our results thus demonstrate the limits of accuracy of the BPM approach, providing guidance for modeling of Raman spectroscopy results and also providing a baseline for the development of more accurate atomistic and neural network  models of polarizability dynamics.

\section{Methods}

\subsection{Bond Polarizability Model and computational methods}
In this work, the total polarizability tensor ($\alpha_{ij}$) is defined as
$$
\alpha_{ij} = \frac{1}{\Omega} \sum_{n} \alpha_{ij}^{n}
$$
where $\Omega$ is the volume and $\alpha_{ij}^{n}$ is the polarizability tensor for the $n^{th}$ bond. Furthermore, $\alpha_{ij}^{n}$ is defined as 
$$
\alpha_{ij}^{n} = \frac{1}{3}(\alpha_{l} + 2\alpha_{p})\delta_{ij} + (\alpha_{l} - \alpha_{p})(\frac{R_{i}R_{j}}{R^{2}} - \frac{1}{3}\delta_{ij}) 
$$
where $i$, $j$ denote $x$, $y$ or $z$ axes. \pmb{$R$} is the vector between two atoms forming the $n^{th}$ bond, and  $\alpha_{l}$ and $\alpha_{p}$ are the longitudinal and perpendicular bond polarizabilities, respectively for a particular bond type.  According to BPM, these parameters are functions of the bond length  $R$ only. Therefore, the bond polarizabilities can be expanded in a Taylor series with respect to the bond length around the equilibrium structure denoted by superscript 0, as given by
$$
\alpha_{t} = \alpha_{t}^{0} + \alpha_{t}^{1}.(R - R^{0}) + \alpha_{t}^{2}.(R - R^{0})^{2} + .. 
$$
where $\alpha_{t}$ is either longitudinal ($t$ = $l$) or perpendicular ($t$ = $p$) component of the bond polarizability. 

In order to extract the unknown parameters of the BPM, we calculated polarizability trajectory using density functional perturbation theory (DFPT)~\cite{DFPT_baroni} as implemented in Quantum ESPRESSO code~\cite{QE}. For all of the systems considered here, $ab$ $initio$ molecular dynamics were  also performed using Quantum ESPRESSO. The
Perdew-Burke-Ernzerhof exchange-correlation functional
for solids (PBEsol) was employed for all systems~\cite{PBEsol}.  
All the unknown parameters in the BPM then were extracted by minimizing the difference between the model results and the  training data set along the DFPT-calculated polarizability trajectory.

For CO$_{2}$, SO$_{2}$, H$_{2}$S, H$_{2}$O and NH$_{3}$,   $ab$ $initio$ molecular dynamics simulations were performed at 300 K. First, we calculated polarizability trajectory directly using DFT for 30, 24, 24, 25 and 24 ps with a time step of 0.0015 ps for the CO$_{2}$, SO$_{2}$, H$_{2}$S, H$_{2}$O and NH$_{3}$ molecules, respectively. However, due to computational cost, we considered the DFT-calculated polarizability trajectory for 23, 24, 23, 20 and 20 ps with a time step of 0.0058, 0.0058, 0.0058, 0.0072 and 0.0072 ps in case of CH$_{4}$, CH$_{2}$O, CH$_{3}$OH, CH$_{3}$CH$_{2}$OH and thiophene, respectively.  We then extracted the BPM parameters for 1$^{st}$- and 2$^{nd}$-order  models   using the DFT-calculated polarizability trajectory data.
\begin{figure*}[ht!]
\centering
\includegraphics[width = 16.0 cm,angle =0]{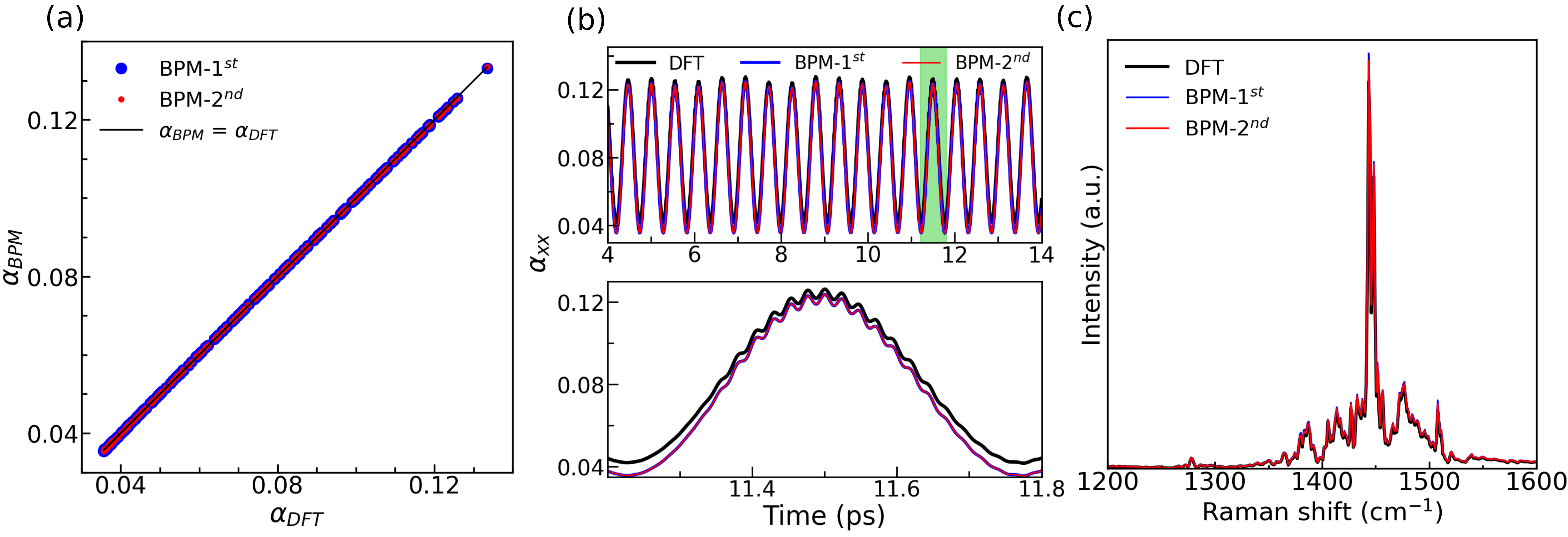}
\caption{(Color online) CO$_{2}$: (a) The $xx$ component of $\alpha_{BPM}$ (1$^{st}$ and 2$^{nd}$ order are shown by  blue and red circles, respectively) vs $\alpha_{DFT}$. The linear fit of BPM-2$^{nd}$ order data is shown by the  black line.. (b) Top panel: Comparison of time-dependent trajectory of $\alpha_{xx}$. Bottom panel: Magnified view of the region in first panel marked in filled green color. (c) Raman spectra from DFT (black), BPM-1$^{st}$ (blue) and BPM-2$^{nd}$ (red).}
\label{Fig1}
\end{figure*}
For BTO and CPB, we found that the polarizability is dominated by the contributions from the motions of the TiO$_{6}$ and PbBr$_{6}$ octahedra (see Supplementary Material (SM)). Therefore, BPM parameters are extracted after fitting to the Ti-O and Pb-Br bonds, respectively, for these systems. We considered a 10-atom unit cell and performed molecular dynamics for 30 ps with a time step of 0.0015 ps while keeping the cell volume fixed at four temperatures of 300 K, 600 K, 800 K and 1000 K, respectively. For the DFT-calculated polarizability trajectory, we  considered the full trajectory of 30 ps with a time step of 0.073 ps. The DFT-calculated polarizability trajectories for all temperatures were used for the extraction of the BPM parameters.

For classical molecular dynamics of BTO, we used a  10$\times$10$\times$10 supercell for 300 ps with a time step of 0.001 ps with the LAMMPS code~\cite{LAMMPS} using the bond-valence model atomistic potential (BVMD)~\cite{BTO_atomistic}. Then,  BPM parameters as extracted from the polarizability calculations for the 10-atom unit cell were used to calculate the Raman spectra of bulk and thin film of BTO for various temperatures. 

\subsection{Experimental details}
Raman scattering was collected in the backscattering configuration ($Z(X,X)\bar{Z}$, X$\parallel$(100)), with a 5 mW 385 nm laser excitation source focused to a surface intensity of 1 W/cm$^2$. Light was dispersed by a triple spectrometer operated in a double-dispersive single-subtractive mode (Horiba Jobin-Yvon, model T64000), and collected using a thermoelectrically-cooled CCD (Andor). The (001)-oriented BaTiO$_3$ bulk single crystal was grown by top seeded solution growth (MTI Corp.).

Raman spectra were collected between 123 K and 423 K at 5 K intervals, and each individual spectrum is the average of 3  acquisitions for 10 s each. Experimental observations of abrupt structural phase transitions at 183 K (rhombohedral (R) to orthorhombic (O)), 278 K (O to tetragonal (T)), and 393 K (T to cubic (C)) are in agreement with the BTO results reported in the literature\cite{C7TA11096K}. Additionally, the intensity of the collapsing soft mode (160 cm$^{-1}$ in the R-phase, and the broad feature at 35 cm$^{-1}$ in the O-phase) are an indication of structural homogeneity and a low population of domain walls, showing that these are effectively single-domain measurements\cite{Yuzyuk2012}.

\section{RESULTS AND DISCUSSION}

\subsection{Linear CO$_2$ molecule}

Fig.~\ref{Fig1} (a) compares the  polarizabilities calculated using BPM and DFT along the polarizability trajectory of the $xx$ component at 300 K for CO$_{2}$. Both 1$^{st}$- and 2$^{nd}$-order BPM reproduce the DFT polarizability very accurately (see the linear fit of the 2$^{nd}$ order BPM data). We also compare polarizability trajectory ($\alpha_{xx}$) as calculated from DFT and BPM (1$^{st}$ and 2$^{nd}$) in Fig.~\ref{Fig1} (b). The top panel of Fig.~\ref{Fig1} (b), which presents the overall polarizability trajectory, shows  perfect agreement between the overall trajectories from BPM and DFT, corresponding to low-frequency vibrations. The bottom panel of Fig.~\ref{Fig1} (b), which presents a magnified view of a small region of the trajectory also shows good agreement between the model and DFT results for the high-frequency polarizability oscillations. This demonstrates the accuracy of BPM on simple linear molecule such as CO$_{2}$. Therefore, it is expected that the Raman spectrum  calculated using BPM will capture the main peaks of the DFT Raman spectrum.  Fig.~\ref{Fig1} (c) shows the perfect overlap of peak positions and intensities of the Raman spectra obtained by these two methods. These results show that the polarizability fluctuations are well-reproduced by the BPM. 
\subsection{Nonlinear molecules}
\begin{figure*}[]
\centering
\includegraphics[width = 16.0 cm,angle =0]{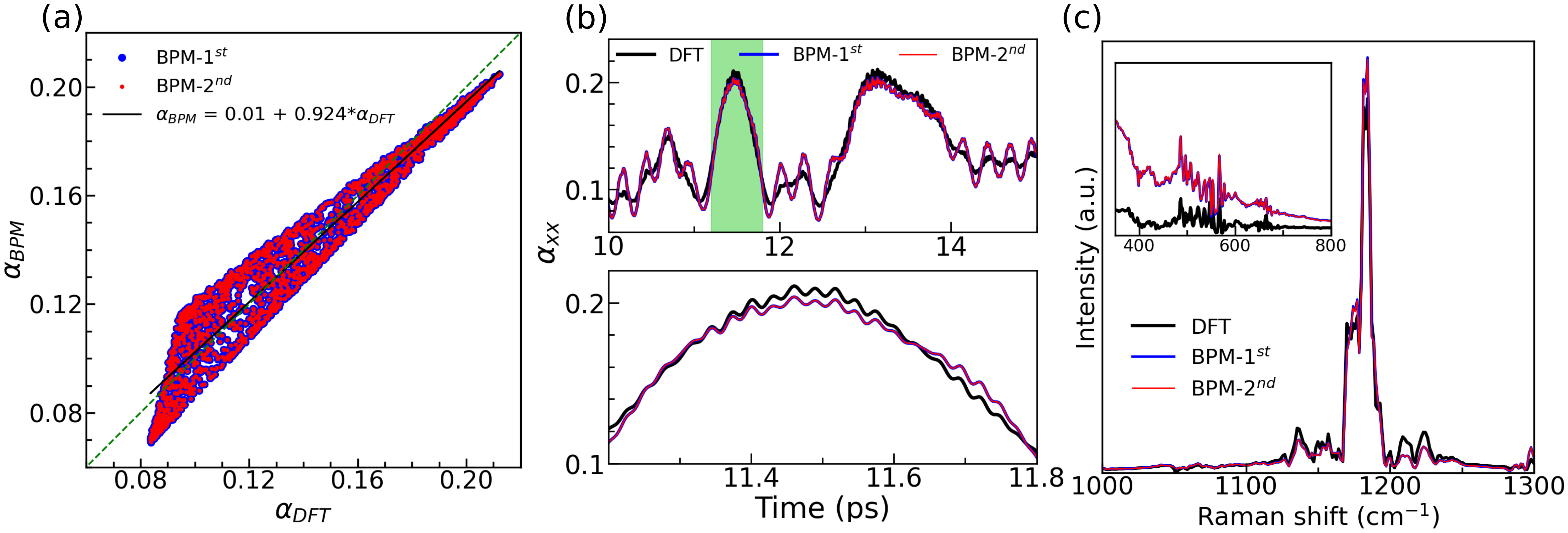}
\caption{(Color online) SO$_{2}$ (a) The $xx$ component of $\alpha_{BPM}$ (1$^{st}$ and 2$^{nd}$ order are shown by blue and red circles, respectively) vs $\alpha_{DFT}$. The linear fit of the BPM-2$^{nd}$ order data is shown by  solid black line. The green dotted line represents  $\alpha_{BPM}$ = $\alpha_{DFT}$. (b) Top panel: Comparison of the trajectory of $\alpha_{xx}$. Bottom panel: Magnified view of the region shaded in green in the  first panel. (c) Raman spectra from DFT (black), BPM-1$^{st}$ (blue) and BPM-2$^{nd}$ (red) results. Inset shows the results in the lower-wavenumber region.}
\label{Fig2} 
\end{figure*}

Then, we considered  the SO$_{2}$, H$_{2}$S, H$_{2}$O, NH$_{3}$ and CH$_{4}$ nonlinear molecules in order  to examine the effect of angular geometry on the BPM accuracy. Fig.~\ref{Fig2} (a) compares the calculated values of the the $xx$ component of $\alpha_{BPM}$ for 1$^{st}$ and 2$^{nd}$ order BPM with $\alpha_{DFT}$ for SO$_{2}$.
The scatter of the data around the $y$ = $x$ line  ($\alpha_{BPM}$ = $\alpha_{DFT}$) shows that the 1$^{st}$-order BPM cannot fully capture the DFT values. 
These results are not improved even after the application of the 2$^{nd}$-order BPM.  
However, an overall linear dependence  of $\alpha_{BPM}$ plotted versus $\alpha_{DFT}$ is observed (the equation of the linear fit of the 2$^{nd}$ order BPM $\alpha$ values plotted versus DFT $\alpha$ values is   $\alpha_{BPM}$ = 0.01 + 0.924*$\alpha_{DFT}$) with an increase of the deviation of the value of $\alpha_{BPM}$ from $\alpha_{DFT}$ when going from larger to smaller values of $\alpha_{DFT}$, forming a triangular-like data distribution. 
We also plotted the polarizability trajectories at 300 K of SO$_{2}$  from these different methods in Fig.~\ref{Fig2} (b). The gross features of the DFT trajectory are  reproduced well by the 1$^{st}$ and 2$^{nd}$ order BPM. Even though the low-frequency (top panel of Fig.~\ref{Fig2} (b)) and high-frequency (bottom panel of Fig.~\ref{Fig2} (b)) polarizability fluctuations are well-captured by the BPM, the amplitude of the low- and high-frequency fluctuations from the BPM results shows some differences from the DFT results. These discrepancies are reflected in the calculated Raman spectra  for the high-frequency region (see Fig.~\ref{Fig2} (c)), where the spectral positions of the DFT-generated spectra are reproduced well by the  BPM-generated spectra. However, unlike for CO$_{2}$, the peak intensities from BPM do not agree  with the DFT results. Additionally, differences between the Raman intensities of the DFT and BPM spectra  are also observed for the low-frequency peaks (see inset Fig.~\ref{Fig2} (c)).         
\begin{figure*}[]
\centering
\includegraphics[width = 16.0 cm,angle =0]{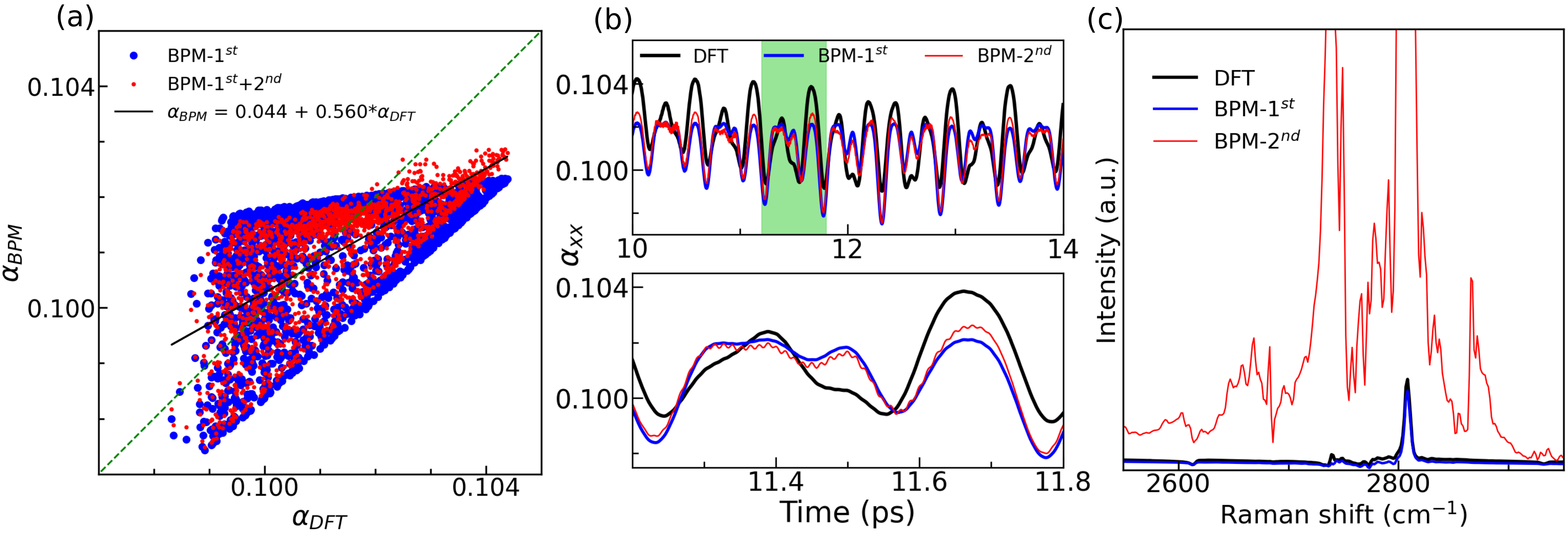}
\caption{(Color online) H$_{2}$S: (Color online) (a) The results for the $xx$ component of $\alpha_{BPM}$ (1$^{st}$ and 2$^{nd}$ order BPM  are shown by blue and red circles, respectively) plotted vs $\alpha_{DFT}$. The linear fit of BPM-2$^{nd}$ order data is shown by the  black line. The green dotted line represents $\alpha_{BPM}$ = $\alpha_{DFT}$. (b) Top panel: Comparison of the trajectory of $\alpha_{xx}$. Bottom panel: Magnified view of  the region in the first panel shaded in green. (c) Raman spectra generated using  DFT (black), BPM-1$^{st}$ (blue) and BPM-2$^{nd}$ (red).}
\label{Fig3}
\end{figure*}

Next, we examined the BPM results for  H$_{2}$S. Fig.~\ref{Fig3} (a), presents the results for BPM-calculated $\alpha$ plotted versus  DFT-calculated $\alpha$ for the $xx$ component. Similar to SO$_{2}$, the 1$^{st}$ order $\alpha_{BPM}$ values show high scatter in comparison to $\alpha_{DFT}$ forming a triangular data distribution where the deviation  of the BPM-generated $\alpha$ values from the  DFT-generated $\alpha$ increases drastically while moving from higher to lower values of DFT-generated $\alpha$. Inclusion of 2$^{nd}$ order in the BPM improves the results slightly so that only a small decrease in the size of the triangular data distribution is obtained. The inclusion of higher-order (3$^{rd}$ and 4$^{th}$ order) terms in the BPM did not eliminate this error, indicating that it is due to the lack of consideration of the effects of the H-S-H angle on the polarizability. The  $\alpha_{xx}$ polarizability trajectories from different methods are shown in Fig.~\ref{Fig3} (b), revealing poor agreement between the BPM- and DFT-calculated polarizability trajectories. A small improvement by the 2$^{nd}$-order model compared to the 1$^{st}$-order model is observed for the overall variation of   $\alpha_{xx}$  (Top panel of Fig.~\ref{Fig3} (b)). However,  a magnified view of the $\alpha_{xx}$ trajectories shows that the 2$^{nd}$-order model trajectory exhibits much stronger high-frequency fluctuations compared to both  the DFT or 1$^{st}$-order model trajectories.  

These much stronger high-frequency fluctuations lead to a dramatically higher Raman intensity for the high-frequency peaks predicted by the 2$^{nd}$-order model compared to the DFT and 1$^{st}$ order model spectra (Fig.~\ref{Fig3} (c)).
Both BPM-1$^{st}$ and BPM-2$^{nd}$ order models predict peak positions similar to those of DFT in the high-frequency spectrum. However, comparison of the obtained peak intensities shows that the BPM-1$^{st}$ spectrum is  closer to the DFT spectrum than the BPM-2$^{nd}$ spectrum.  The unphysicallly high intensity of the high-frequency fluctuations in the BPM-2$^{nd}$ order model is likely due to the fact that the model parameterization procedure seeks to minimize the overall error; since the high-frequency oscillations are much weaker than the low-frequency oscillations, an overestimation of high-frequency amplitude will be acceptable in the parameterization if it is balanced by better agreement with the overall trajectory. However, this will lead to unphysically high intensity in the physically relevant part of the spectrum.

\begin{figure*}[]
\centering
\includegraphics[width = 16.0 cm,angle =0]{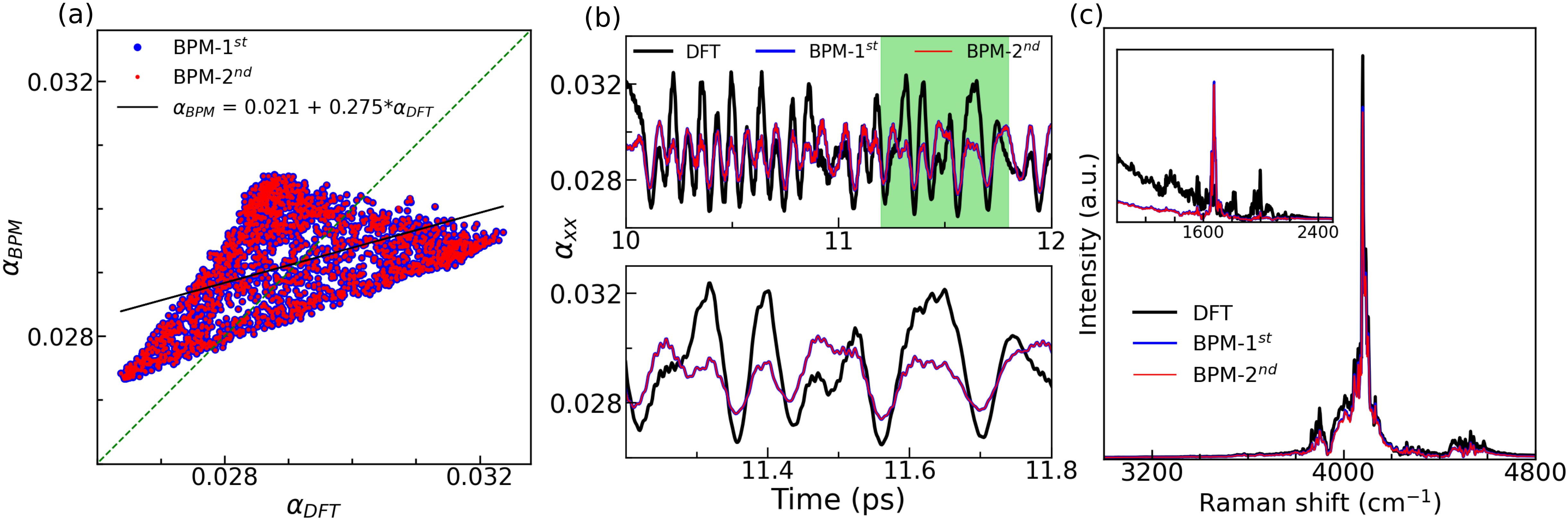}
\caption{(Color online) H$_{2}$O: (Color online) (a) The results for the $xx$ component of $\alpha_{BPM}$ (1$^{st}$ and 2$^{nd}$ order BPM  are shown by blue and red circles, respectively) plotted vs $\alpha_{DFT}$. The linear fit of BPM-2$^{nd}$ order data is shown by the  black line. The green dotted line represents $\alpha_{BPM}$ = $\alpha_{DFT}$. (b) Top panel: Comparison of the trajectory of $\alpha_{xx}$. Bottom panel: Magnified view of  the region in the first panel shaded in green. (c) Raman spectra generated using  DFT (black), BPM-1$^{st}$ (blue) and BPM-2$^{nd}$ (red).}
\label{Fig4}
\end{figure*}

We then examine the performance of BPM for  H$_{2}$O, perhaps the most important molecule in chemistry and biology.  Fig.~\ref{Fig4} (a) shows $\alpha_{xx}$  calculated using BPM-1$^{st}$ and BPM-2$^{nd}$  for different  H$_{2}$O geometries plotted versus the corresponding DFT $\alpha_{xx}$ values.
Similar to other nonlinear molecules, a triangular data distribution is observed for both BPM-1$^{st}$ and BPM-2$^{nd}$. However, unlike for H$_{2}$S, the inclusion of 2$^{nd}$ order does not improve the BPM results. The calculated slope (0.275) of the linear fit of the BPM-2$^{nd}$  vs DFT data is even worse than  that for H$_{2}$S (0.560). We also plot the polarizability trajectories as calculated from BPM-1$^{st}$, BPM-2$^{nd}$ and DFT methods in  Fig.~\ref{Fig4} (b). The low-frequency polarizability trajectory (see top panel of Fig.~\ref{Fig4} (b)) is not well-captured by BPM. However, the high-frequency oscillations of the polarizability trajectory are well-reproduced by  BPM (see bottom panel of Fig.~\ref{Fig4} (b)). This  implies that the BPM-calculated Raman spectrum will be in good agreement with the DFT spectrum in the high-frequency region but not in the low-frequency region. The  Raman spectra (Fig.~\ref{Fig4} (c)) calculated from the polarizability trajectory show that the peak position in the high-frequency range is well-reproduced by the BPM-1$^{st}$ and BPM-2$^{nd}$. However, a small disagreement between the peak  intensities between BPM and DFT spectra is observed in this region. By contrast, the spectra in the low-frequency region (inset of Fig.~\ref{Fig4} (c)) do not show  good agreement with the DFT results.     

\begin{figure*}[]
\centering
\includegraphics[width = 16.0 cm,angle =0]{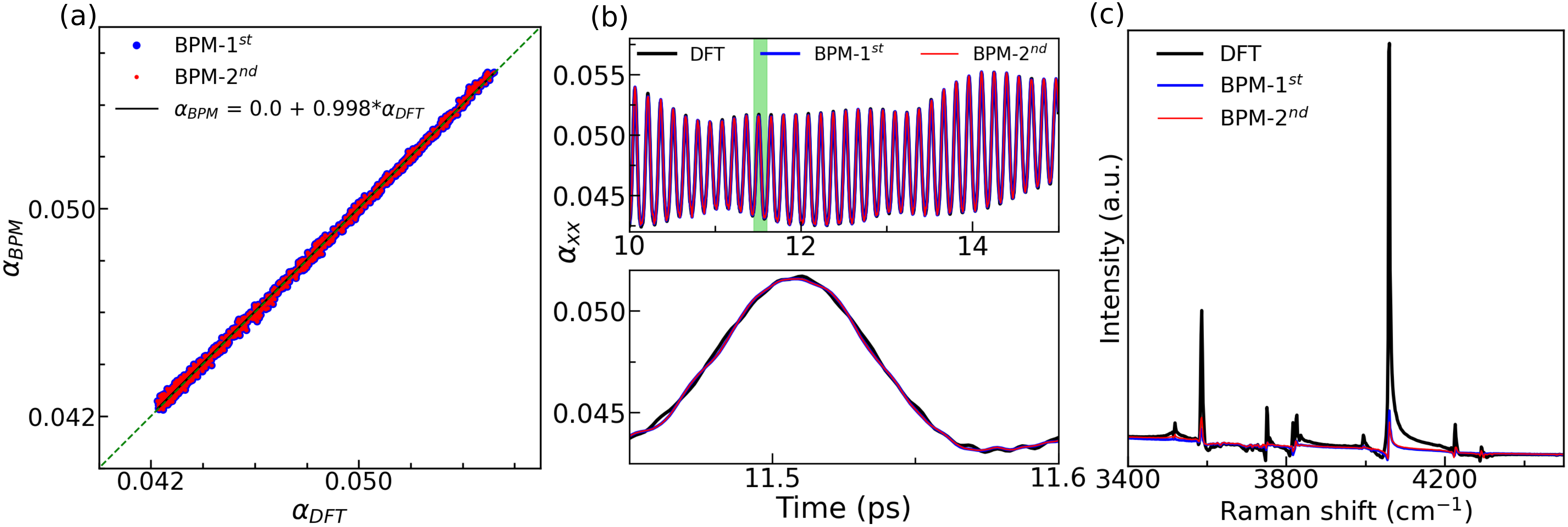}
\caption{(Color online) NH$_{3}$: (Color online) (a) The results for the $xx$ component of $\alpha_{BPM}$ (1$^{st}$ and 2$^{nd}$ order BPM  are shown by blue and red circles, respectively) plotted vs $\alpha_{DFT}$. The linear fit of BPM-2$^{nd}$ order data is shown by the  black line. The green dotted line represents $\alpha_{BPM}$ = $\alpha_{DFT}$. (b) Top panel: Comparison of the trajectory of $\alpha_{xx}$. Bottom Panel: Magnified view of the region in the top panel shaded in green. (c) Raman spectra generated using  DFT (black), BPM-1$^{st}$ (blue) and BPM-2$^{nd}$ (red).}
\label{Fig5}
\end{figure*}

Next, we consider  NH$_{3}$ and CH$_{4}$ which are nonlinear molecules with multiple equal bond angles.  While naively,  one may expect the BPM polarizability to show stronger disagreement with DFT results for  NH$_{3}$ compared to H$_{2}$O due to a greater number of bond angles, in contrast to the results for H$_{2}$S, H$_{2}$O and SO$_{2}$, the plot of BPM polarizability versus DFT polarizability shows  a good linear trend Fig.~\ref{Fig5} (a). The overall $\alpha_{xx}$ trajectory is also reproduced well (top panel of Fig.~\ref{Fig5} (b)). However, a magnified view of the trajectory shows significant differences between the high-frequency $\alpha_{xx}$ oscillations predicted by BPM and those obtained by DFT calculations, with much larger amplitude of oscillations for the DFT trajectory (bottom panel of Fig.~\ref{Fig5} (b)).   Correspondingly, examination of the Raman spectra presented in Fig.~\ref{Fig5} (c) shows that the high-frequency peaks of the DFT spectrum have much higher intensity than the corresponding BPM peaks. While BPM and DFT obtain the same frequencies for the peaks, the relative  intensities of the peaks  are somewhat different.  Thus, similar to  H$_{2}$S, H$_{2}$O and SO$_{2}$, BPM correctly reproduces the peak positions but does not accurately reproduce the fine features of the DFT spectra. 
\begin{figure*}[]
\centering
\includegraphics[width = 16.0 cm,angle =0]{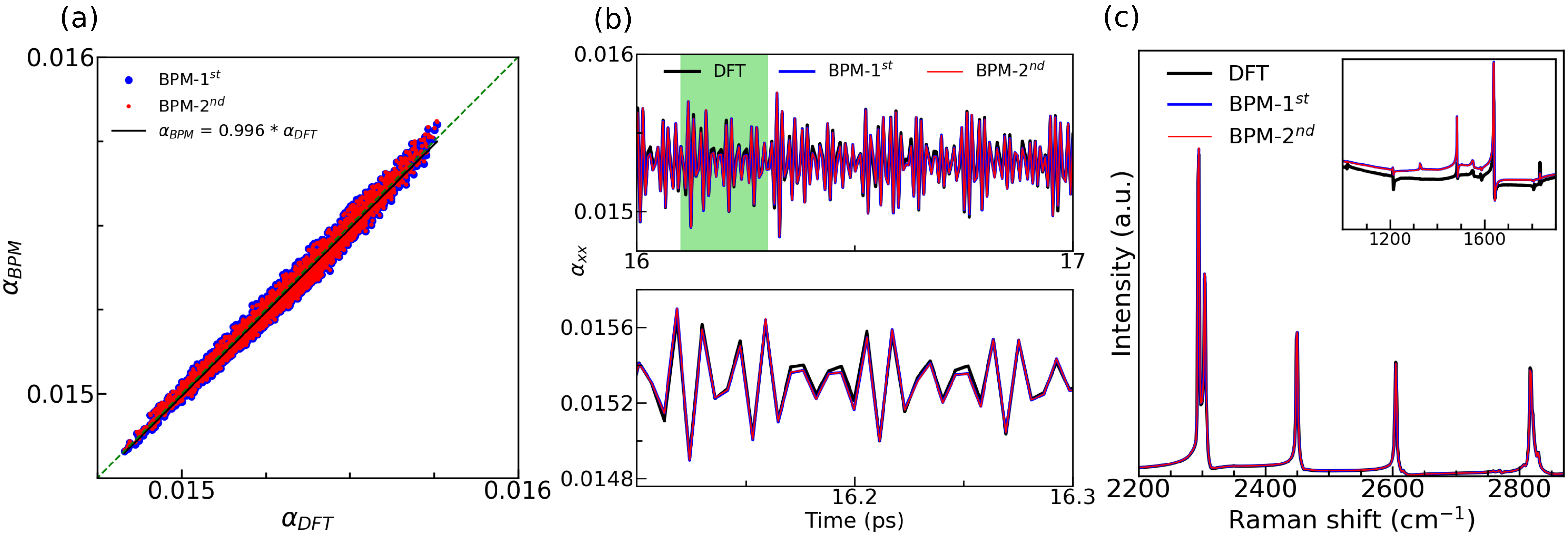}
\caption{(Color online) CH$_{4}$: (Color online)  (a) The results for the $xx$ component of $\alpha_{BPM}$ (1$^{st}$ and 2$^{nd}$ order BPM  are shown by blue and red circles, respectively) plotted vs $\alpha_{DFT}$. The linear fit of BPM-2$^{nd}$ order data is shown by the  black line. The green dotted line represents $\alpha_{BPM}$ = $\alpha_{DFT}$. (b) Top panel: Comparison of the trajectory of $\alpha_{xx}$. Bottom panel: Magnified view of  the region in the first panel shaded in green. (c) Raman spectra generated using  DFT (black), BPM-1$^{st}$ (blue) and BPM-2$^{nd}$ (red).}
\label{Fig6}
\end{figure*}

Next, examination of the CH$_{4}$ results in Fig.~\ref{Fig6} reveals good  agreement between BPM and DFT results with only a small deviation from $y$ = $x$ line.  A close examination of the polarizability trajectory and the Raman spectrum reveals small but noticeable differences. Thus, while accurate, the BPM for CH$_{4}$ does not show perfect agreement such as that found for CO$_{2}$, indicating that the lack of a bond-angle-dependent term  in the BPM does affect the accuracy of model polarizability, even though the effect is weak.

\begin{figure*}[]
\centering
\includegraphics[width = 16.0 cm,angle =0]{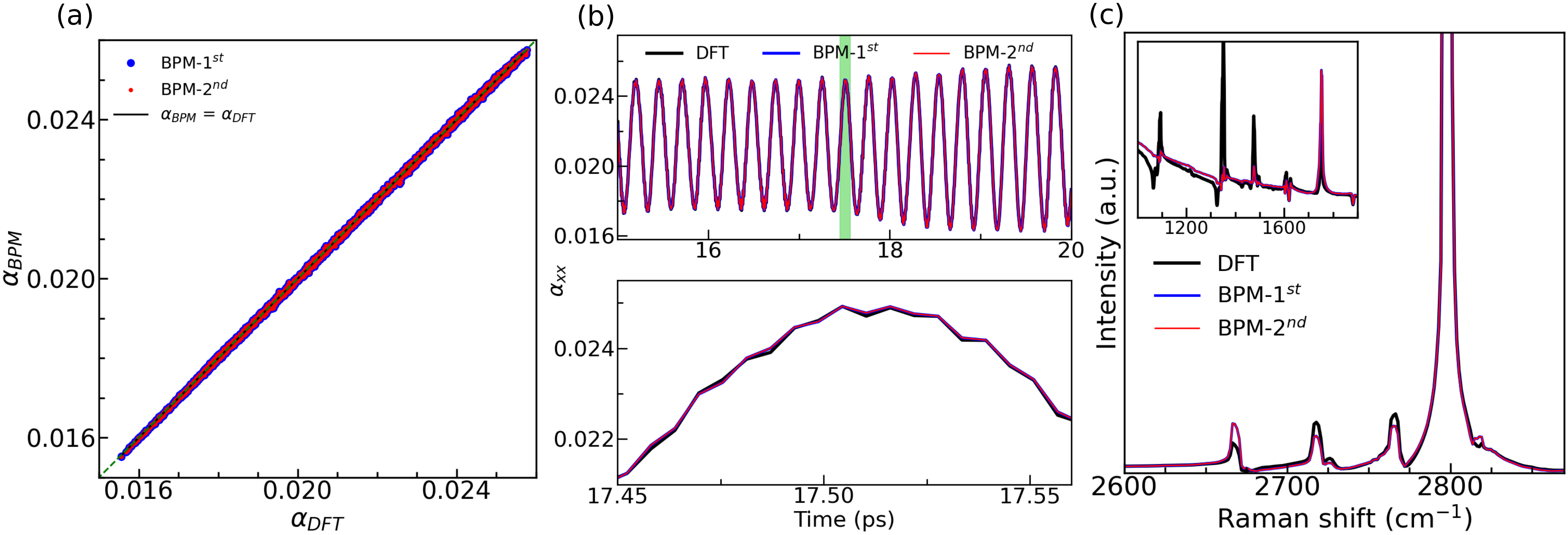}
\caption{(Color online) CH$_{2}$O: (Color online) (a) The results for the $xx$ component of $\alpha_{BPM}$ (1$^{st}$ and 2$^{nd}$ order BPM  are shown by blue and red circles, respectively) plotted vs $\alpha_{DFT}$. The linear fit of BPM-2$^{nd}$ order data is shown by the  black line. The green dotted line represents $\alpha_{BPM}$ = $\alpha_{DFT}$. (b) Top panel: Comparison of the trajectory of $\alpha_{xx}$. Bottom panel: Magnified view of  the region in the first panel shaded in green. (c) Raman spectra generated using  DFT (black), BPM-1$^{st}$ (blue) and BPM-2$^{nd}$ (red).}
\label{Fig7}
\end{figure*}

We now discuss the origin of the polarizability errors of the BPM in the angular molecules.   The BPM is a Taylor expansion of the polarizability around the ground state structure, where rather than carrying out the Taylor expansion in terms of the Cartesian coordinates of the atoms, the Taylor expansion is carried out in terms of the deviations of the bond lengths from their ground state values.  However, for tri-atomic molecules, in addition to the bond lengths, the bond angle can also affect the polarizability and a full Taylor expansion should  involve angle-dependent  terms such a  $a_\theta \Delta \theta$ + $b_\theta (\Delta \theta)^2$ +.. as well as cross-terms describing the interplay of the effects of the bond length and bond angle, such as $a_{r,\theta} \Delta \theta \Delta r$  + ...  . Clearly, these terms are absent in the BPM that only takes into account the changes in the bond lengths and ignores the effect of the angle. 

For a Taylor expansion, we expect that higher-order terms are weaker than the lower-order terms.  Thus, the linear effect of the change in $\theta$ given by $a_{\theta} \Delta \theta$ is larger than the quadratic effect given by $b_{\theta} (\Delta \theta)^2$.  In a tri-atomic angular molecule such as water, a vibration that changes the bond angle will change the polarizability by $a_{\theta} \Delta \theta$ + $b_{\theta} \Delta \theta^2$ + $a_{r,\theta} \Delta \theta \Delta r$. For the water molecule with  only one bond angle, any motion of the O-H bonds that changes the bond angle is uncompensated by an equal and opposite change in a different bond angle, so that the error is of the order $a_{\theta} \Delta \theta$ + $b_{\theta} (\Delta \theta)^2$ + $a_{r,\theta} \Delta \theta \Delta r$.  However, for a molecule such as CH$_{4}$, any movement of the C-H bond that leads to a change in the bond angle makes one bond angle smaller and another bond angle greater. This leads to the cancellation of the first-order effects of both the $\theta$  and the $\theta r$ terms, so that the error is given by  $b_{\theta} (\Delta \theta)^2$ only  and is much smaller than that for water.   For NH$_{3}$, the size of the error depends on the direction of vibration. For a vibration by an N-H bond toward another N-H bonds, the linear effects will be canceled out and the error due to ignoring the angular effect will be only second order in $\Delta \theta$ and  small. However for a vibration by an N-H bond toward the N lone pair, the linear angular effect will be present and the error will be large.  Therefore, we observe very larger deviation of BPM results for polarizability in the $\alpha_{BPM}$ vs $\alpha_{DFT}$ plot for angular tri-atomic molecules H$_{2}$O, H$_{2}$S and SO$_{2}$, a much smaller deviation for NH$_{3}$ and an even  smaller deviation for CH$_{4}$. 

\begin{figure*}[]
\centering
\includegraphics[width = 16.0 cm,angle =0]{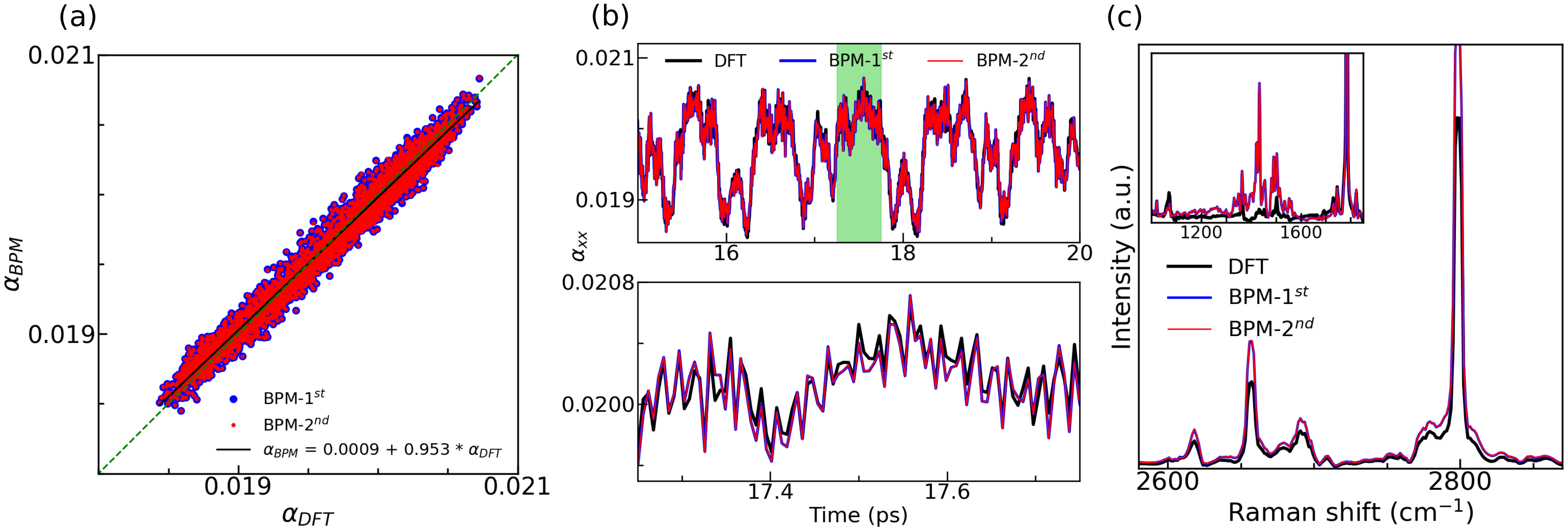}
\caption{(Color online) CH$_{3}$OH: (Color online) (a) The results for the $xx$ component of $\alpha_{BPM}$ (1$^{st}$ and 2$^{nd}$ order BPM  are shown by blue and red circles, respectively) plotted vs $\alpha_{DFT}$. The linear fit of BPM-2$^{nd}$ order data is shown by the  black line. The green dotted line represents $\alpha_{BPM}$ = $\alpha_{DFT}$. (b) Top panel: Comparison of the trajectory of $\alpha_{xx}$. Bottom panel: Magnified view of  the region in the first panel shaded in green. (c) Raman spectra generated using  DFT (black), BPM-1$^{st}$ (blue) and BPM-2$^{nd}$ (red).}
\label{Fig8}
\end{figure*}

\begin{figure*}[]
\centering
\includegraphics[width = 16.0 cm,angle =0]{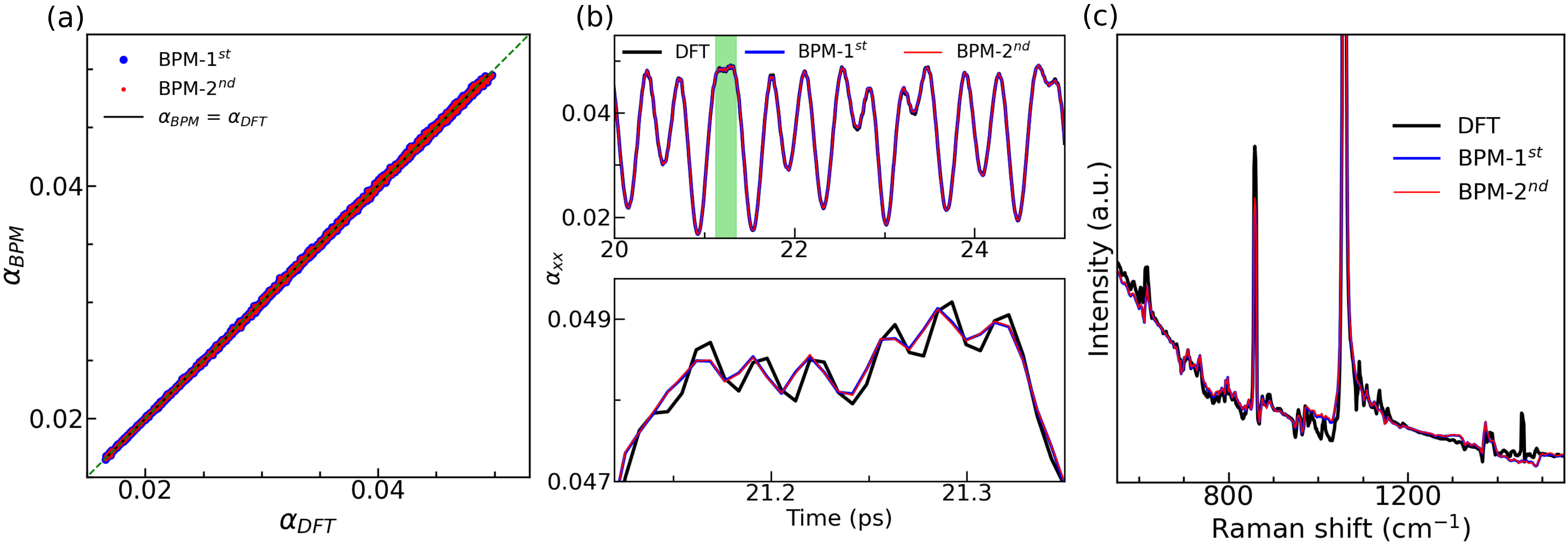}
\caption{(Color online) Thiophene: (Color online)  (a) The results for the $xx$ component of $\alpha_{BPM}$ (1$^{st}$ and 2$^{nd}$ order BPM  are shown by blue and red circles, respectively) plotted vs $\alpha_{DFT}$. The linear fit of BPM-2$^{nd}$ order data is shown by the  black line. The green dotted line represents $\alpha_{BPM}$ = $\alpha_{DFT}$. (b) Top panel: comparison of the trajectory of $\alpha_{xx}$. Bottom panel: Magnified view of  the region in the first panel shaded in green. (c) Raman spectra generated using  DFT (black), BPM-1$^{st}$ (blue) and BPM-2$^{nd}$ (red).}
\label{Fig9}
\end{figure*}

\begin{figure*}[]
\centering
\includegraphics[width = 16.0 cm,angle =0]{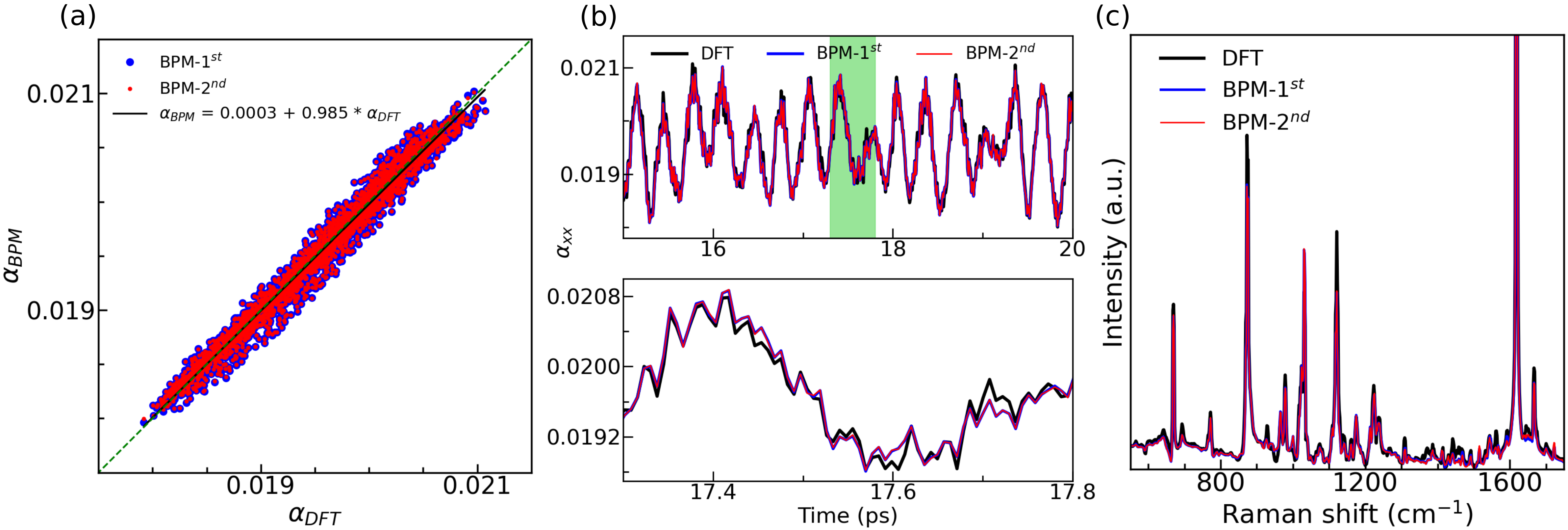}
\caption{(Color online) CH$_{3}$CH$_{2}$OH: (Color online)  (a) The results for the $xx$ component of $\alpha_{BPM}$ (1$^{st}$ and 2$^{nd}$ order BPM  are shown by blue and red circles, respectively) plotted vs $\alpha_{DFT}$. The linear fit of BPM-2$^{nd}$ order data is shown by the  black line. The green dotted line represents $\alpha_{BPM}$ = $\alpha_{DFT}$. (b) Top panel: Comparison of the trajectory of $\alpha_{xx}$. Bottom panel: Magnified view of  the region in the first panel shaded in green. (c) Raman spectra generated using  DFT (black), BPM-1$^{st}$ (blue) and BPM-2$^{nd}$ (red).}
\label{Fig10}
\end{figure*}

To further test the performance of BPM, we examine the more complicated tri-element molecules CH$_{2}$O (see Fig.~\ref{Fig7}) and CH$_{3}$OH (see Fig.~\ref{Fig8}). We find that despite the strong asymmetry of CH$_{2}$O, BMP shows small deviations from the DFT results.  This is likely due to the high stiffness of the 120$^{\circ}$ H-C-O and the H-C-H  angles that only allow small vibrations (considerably smaller than those for water) and therefore lead to only small deviations of BPM polarizability from the DFT polarizability even in the presence of error linear in $\theta$.  For CH$_{3}$OH, we find significant differences between  BPM  and DFT polarizability values, due to the larger fluctuations of the C-O-H bond angle for which the terms linear in $\theta$ are not canceled out.  These results suggest that BPM will be accurate in the case where linear dependence of the polarizability on the bond angle is either zero due to the presence of multiple bonds such that the bond motion makes one bond angle greater and the other bond angle smaller, or is small due to the stiff double  bonds.  On the other hand, for highly asymmetric molecules with single bonds, and in particular for O atoms  where only a single bond angle centered at O is possible, BPM errors due to ignoring the effect of $\theta$  will be significant. Comparison of the BPM results for polarizability and Raman spectra of thiophene (C$_{4}$H$_{4}$S) and  CH$_{3}$CH$_{2}$OH with the corresponding DFT results confirm the above conclusions.  For the aromatic thiophene molecule, small errors are obtained for the BPM polarizability results as shown in Figs.~\ref{Fig9}, with good agreement obtained for the Raman spectra except for the peak at 1450 cm$^{-1}$.  This is due to the stiffness of the aromatic structure with 120$^{\circ}$ bond angles, similar to CH$_2$O.  For CH$_{3}$CH$_{2}$OH, the agreement between the BPM and DFT polarizability values is similar to that for CH$_{3}$OH (see Figs.~\ref{Fig10}) and the Raman spectrum shows a similar poor agreement with the DFT spectrum for the 1200-1600~cm$^{-1}$ region. These results suggest that spectral features related to asymmetric bonds will be reproduced relatively poorly by BPM for large molecule, even while the overall spectrum will show good agreement with DFT results due to the dominance of the functional groups for which the BPM model performs well (e.g. CH$_{3}$ and CH$_{2}$ in ethanol).

\begin{figure*}[]
\centering
\includegraphics[width = 16.0 cm,angle =0]{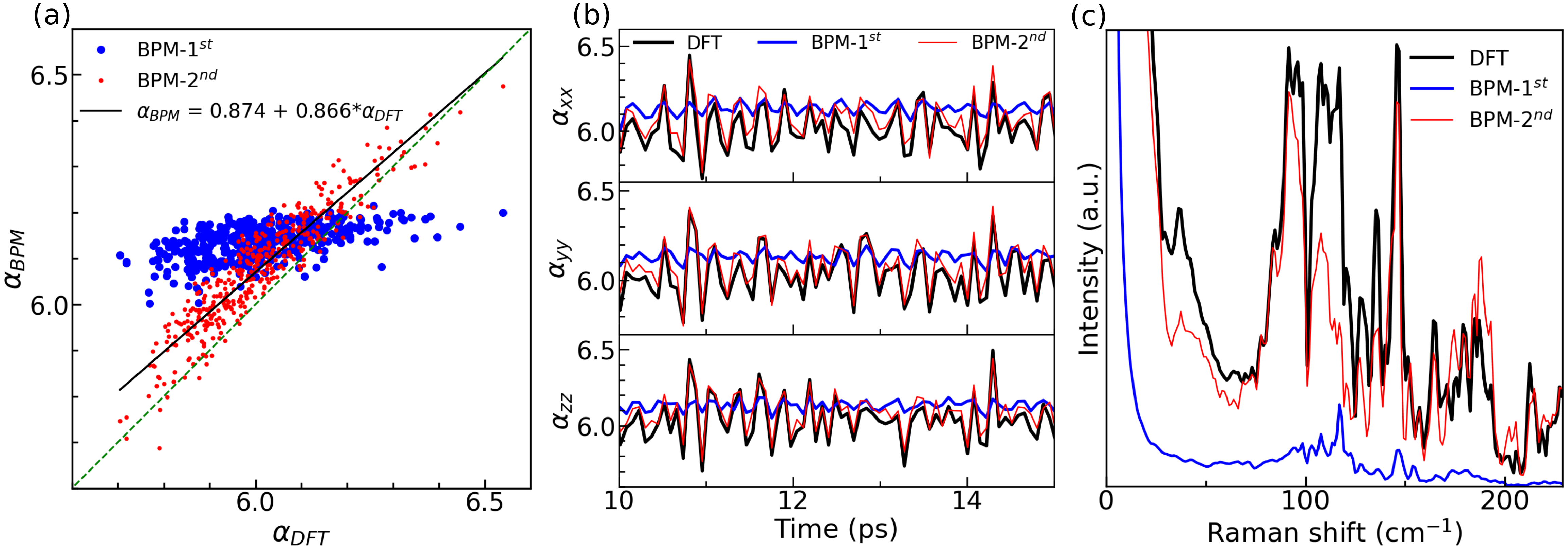}
\caption{(Color online) BaTiO$_{3}$ at 300 K: (Color online) (a) The results for the $xx$ component of $\alpha_{BPM}$ (1$^{st}$ and 2$^{nd}$ order BPM  are shown by blue and red circles, respectively) plotted vs $\alpha_{DFT}$. The linear fit of BPM-2$^{nd}$ order data is shown by the  black line. The green dotted line represents $\alpha_{BPM}$ = $\alpha_{DFT}$. (b) Comparison of the trajectory of $\alpha_{xx}$ (top panel), $\alpha_{yy}$ (middle panel) and $\alpha_{zz}$ (bottom panel) calculated from DFT (black), BPM-1$^{st}$ (blue) and BPM-2$^{nd}$ (red), (c) Raman spectra generated using  DFT (black), BPM-1$^{st}$ (blue) and BPM-2$^{nd}$ (red).}
\label{Fig11}
\end{figure*}

\begin{figure*}[]
\centering
\includegraphics[width = 16.0 cm,angle =0]{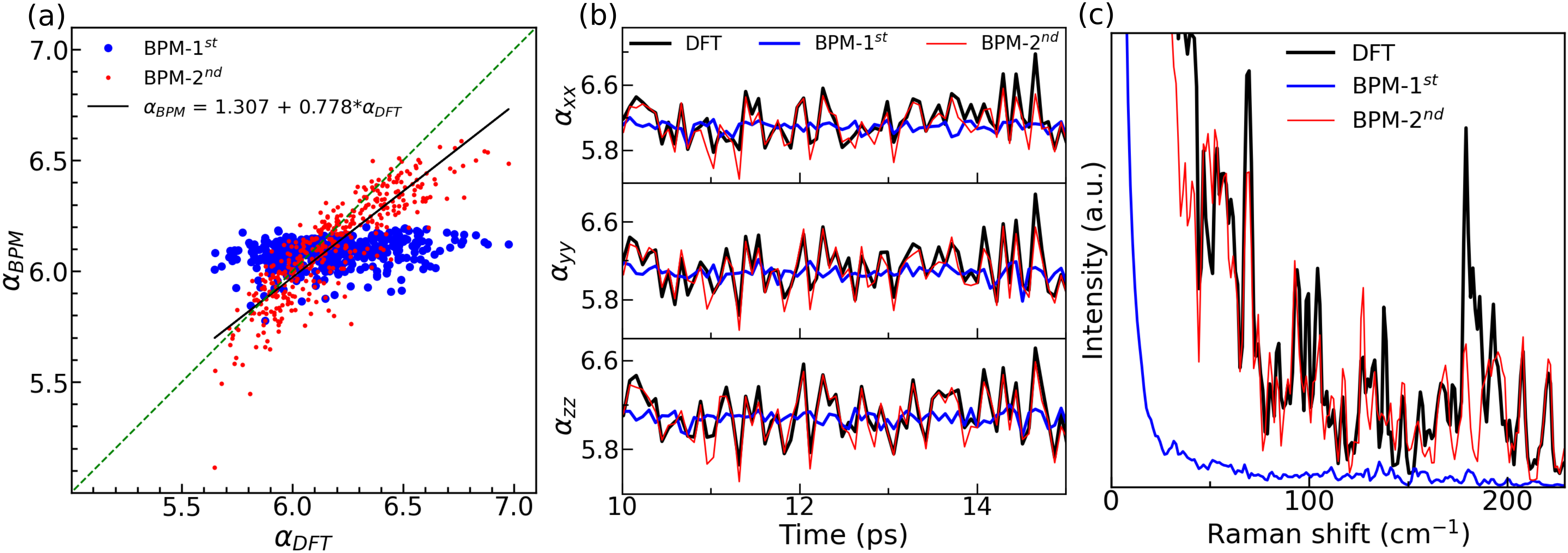}
\caption{(Color online) BaTiO$_{3}$ at 1000 K: (Color online) (a) The results for the $xx$ component of $\alpha_{BPM}$ (1$^{st}$ and 2$^{nd}$ order BPM  are shown by blue and red circles, respectively) plotted vs $\alpha_{DFT}$. The linear fit of BPM-2$^{nd}$ order data is shown by the  black line. The green dotted line represents $\alpha_{BPM}$ = $\alpha_{DFT}$. (b) Comparison of the trajectory of $\alpha_{xx}$ (top panel), $\alpha_{yy}$ (middle panel) and $\alpha_{zz}$ (bottom panel) calculated from DFT (black), BPM-1$^{st}$ (blue) and BPM-2$^{nd}$ (red), (c) Raman spectra generated using  DFT (black), BPM-1$^{st}$ (blue) and BPM-2$^{nd}$ (red).}
\label{Fig12}
\end{figure*}

\subsection{Perovskites}

Multicenter bonding is important for solid-state materials and we therefore examine two perovskites, the classic  ferroelectric BTO and the CPB halide perovskite, which are  representative materials of the important  oxide and halide perovskites in order 
to  understand the effectiveness of the BPM for representing solid-state polarizability and Raman spectra. The perovskite structure enables a variety of different distortions that provide a good test of atomistic polarizability models. For example, BTO can assume four different  phases generated by different directions of average Ti displacements from the center of the O$_6$ octahedra that  show different Raman spectra.  Additionally,  the presence of vibrational modes related to O$_6$ rotations also contributes to the Raman spectra of BTO.    The soft bonds and strong  distortions of  CPB  make it a particularly stringent test case for atomistic polarizability models~\cite{berger2023polarizability}.   

\begin{figure*}[]
\centering
\includegraphics[width = 16.0 cm,angle =0]{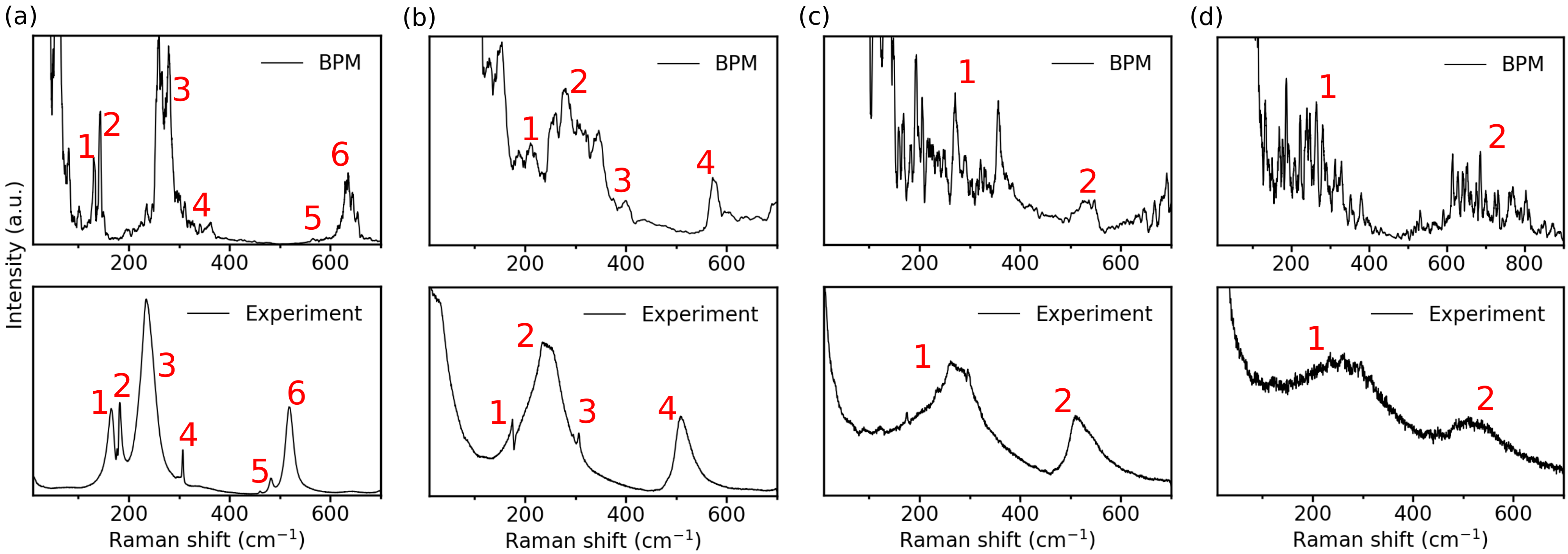}
\caption{(Color online) Raman spectra of different phases ((a) rhombohedral, (b) orthorhombic), (c) tetragonal, and (d) cubic) of BTO single crystal from bond polarizability model (top panel) and experiment (bottom panel). The temperatures for (a), (b), (c) and (d) are 60 K, 88 K, 110 K and 160 K, respectively (top panel), and 123 K, 233 K, 333 K, 423 K, respectively (bottom panel). Peaks from BPM and experiment are marked by number for comparison.}
\label{Fig13}
\end{figure*}

Fig.~\ref{Fig11} (a) plots the $\alpha_{xx}$ values for both 1$^{st}$ and 2$^{nd}$ order BPM versus the DFT-calculated polarizability  for ten-atom supercell of BTO at 300 K. It is observed that the BPM-1$^{st}$ order completely fails to reproduce the DFT-calculated polarizability data, whereas the inclusion of 2$^{nd}$ order term in the BPM strongly  improves the model predictions.  For example, the slope  of the line fit to the BPM-2 data plotted versus DFT data is  0.874, which is relatively close to the $\alpha_{BPM}$ = $\alpha_{DFT}$ straight line. Nevertheless, large scatter is observed in the plot of DFT versus BPM-2$^{nd}$ polarizability.  Inclusion of higher-order dependence of bond polarizability on bond length does not strongly improve the agreement of BPM with DFT (see SM), suggesting that the BPM error is due to the fundamental approximation of independent bond contributions to the polarizability of the BPM.

We then compare the polarizability trajectories ($\alpha_{xx}$, $\alpha_{yy}$, $\alpha_{zz}$) calculated from BPM (1$^{st}$ and 2$^{nd}$) and DFT methods as shown in Fig.~\ref{Fig11} (b). The improvement of BPM results can be observed clearly after the inclusion of BPM-2$^{nd}$ order, which reproduces the overall trajectory well but shows local deviations from the DFT values. Here, we only compare the low-frequency region due to the large time step used to  evaluate the polarizability trajectory. 

The  Raman spectra calculated using the polarizability trajectory with large time steps are shown in Fig.~\ref{Fig11} (c). 
The  BPM-2$^{nd}$ spectrum reproduces all of the peaks of the DFT spectrum, but shows strong differences in intensity for four peaks (at 110, 130, 170 and 215 cm$^{-1}$), indicating that BPM-2$^{nd}$ cannot reliably reproduce the lineshapes of the peaks in the Raman spectrum.

We then applied the same BPM parameters for the structures obtained in the DFT MD simulation of BTO  performed at 1000~K  in order to check the applicability of the BPM parameters at higher temperatures. Fig.~\ref{Fig12} (a) shows the comparison of the BPM-calculated polarizability with the DFT-calculated polarizability. Similar to the 300~K results (Fig.~\ref{Fig11} (a)), the BPM-1$^{st}$ order fail to reproduce the DFT-results. However, the BPM-2$^{nd}$ improved the value of polarizability so that the data points are more aligned along the $\alpha_{BPM}$ = $\alpha_{DFT}$ straight line. The  scatter of the data is greater than that for 300 K, most likely due to the more anharmonic vibrations, greater impact of the changes in the Ti-O-Ti and O-Ti-O bond angles, and greater coupling between the bonds. 
This can also be seen by comparing the calculated slope from the linear fitting, where the calculated value of the slopes are 0.866 and 0.778 at 300 K and 1000 K, respectively. 

The direct comparison of the polarizability trajectories at 1000 K is shown in Fig.~\ref{Fig12} (b). Similar to the results for the simulation at 300 K, the BPM-2$^{nd}$ captures the overall fluctuation of the polarizability  very well. These results are also reflected in the calculated Raman spectra plot for the low-frequency region  as shown in Fig.~\ref{Fig12} (c).  Interestingly, despite the stronger deviation of the BPM-2$^{nd}$ polarizabilities from DFT polarizability for 1000 K than 300 K, the Raman spectrum at 1000 K shows better agreement between DFT and BPM-2$^{nd}$ with only the peak at 180 cm$^{-1}$ showing strong underestimation of the intensity by BPM-2$^{nd}$ compared to DFT.  This is likely either due to error cancellation or because the high errors of BPM-2$^{nd}$ are relevant for non-Raman-active modes that do not contribute  to the error in the Raman spectrum. The lack of reliability of BPM-2$^{nd}$ for the peak intensity means that the temperature evolution of the Raman peaks is reproduced well for some peaks but is only qualitatively reproduced for other peaks. For example, DFT results for the ten-atom cell dynamics show that the intensity of the peak at 180 cm$^{-1}$ is weaker than that at 100 cm$^{-1}$ at 300 K and then increases with higher temperature  so that at 1000 K the  peak at 180 cm$^{-1}$ has much higher  intensity  than the peak at 100 cm$^{-1}$. By contrast, BPM-2$^{nd}$ spectra show that the peak at 180 cm$^{-1}$ changes from weaker intensity to same intensity as the peak at 100 cm$^{-1}$ when the temperature is changed from  300 to 1000 K.  Thus, while the BPM qualitatively reproduces the overall trend of the changes in the relative intensities of the two peaks, it does not  provide quantitative agreement with DFT.

\begin{figure}[]
\centering
\includegraphics[width = 8.5 cm,angle =0]{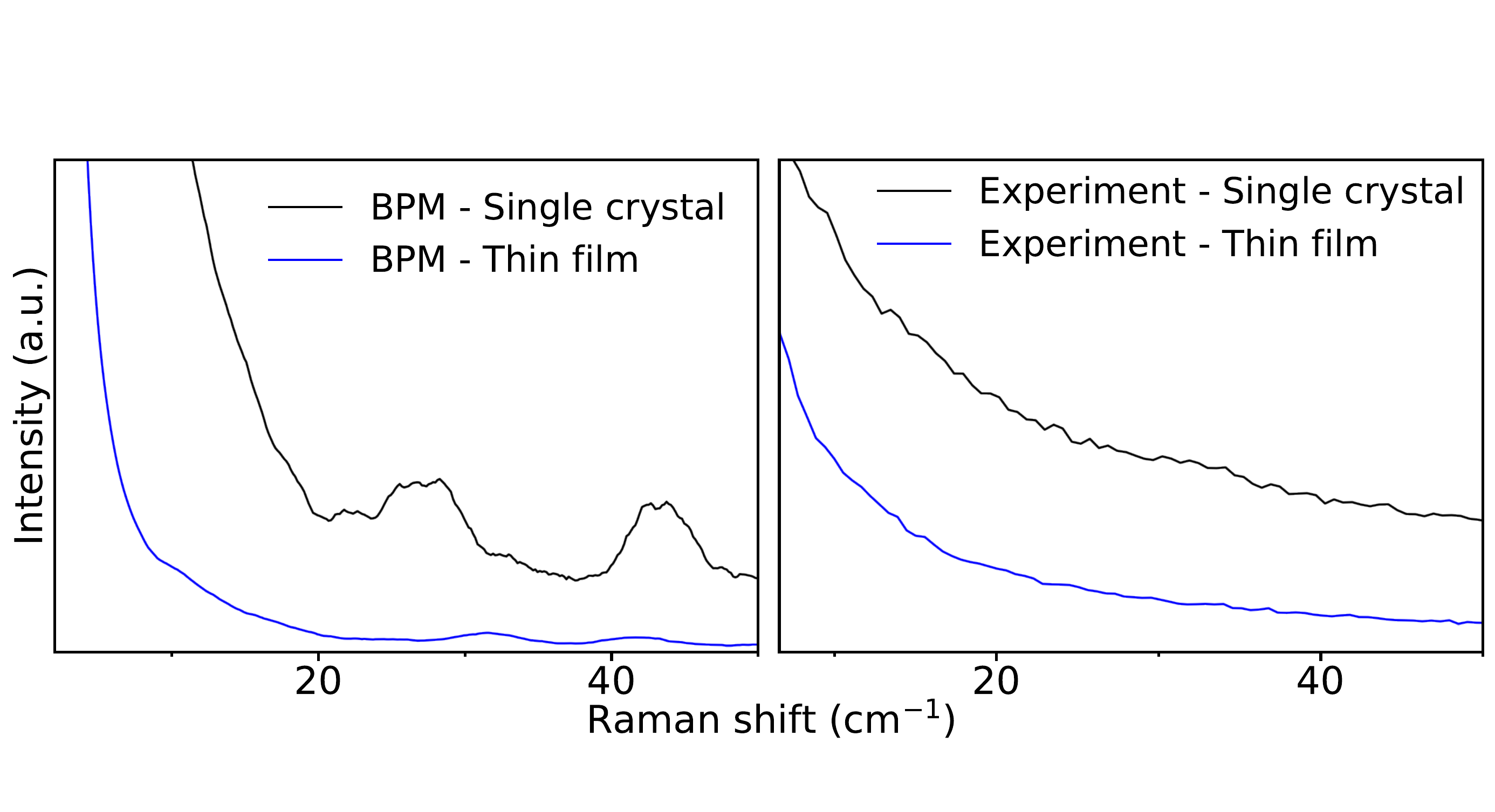}
\caption{(Color online) Central peak intensity of single crystal (black) and thin film (blue) of BTO (tetragonal phase) from BPM (left panel) and experiment (right panel). The temperature for left panel and right panel are 130 K and 333 K, respectively.}
\label{Fig14}
\end{figure}
\begin{figure*}[]
\centering
\includegraphics[width = 16.0 cm,angle =0]{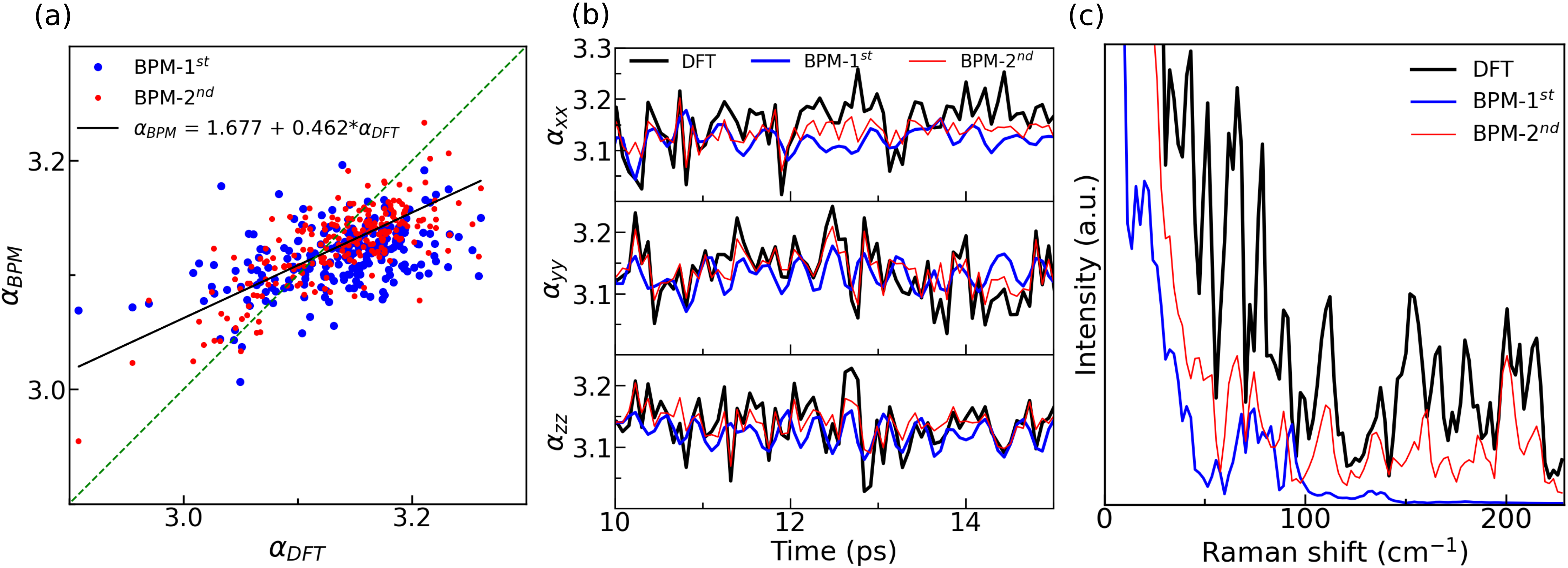}
\caption{(Color online) CsPbBr$_{3}$ at 300 K: (Color online) (a) The results for the $xx$ component of $\alpha_{BPM}$ (1$^{st}$ and 2$^{nd}$ order BPM  are shown by blue and red circles, respectively) plotted vs $\alpha_{DFT}$. The linear fit of BPM-2$^{nd}$ order data is shown by the  black line. The green dotted line represents $\alpha_{BPM}$ = $\alpha_{DFT}$. (b) Comparison of the trajectory of $\alpha_{xx}$ (top panel), $\alpha_{yy}$ (middle panel) and $\alpha_{zz}$ (bottom panel) calculated from DFT (black), BPM-1$^{st}$ (blue) and BPM-2$^{nd}$ (red), (c) Raman spectra generated using  DFT (black), BPM-1$^{st}$ (blue) and BPM-2$^{nd}$ (red).}
\label{Fig15}
\end{figure*}

The above BTO results  are calculated for a large time step and a short trajectory because of the large computational cost of DFT-calculated polarizability. Therefore, to examine the performance of the BPM for obtaining Raman spectra of realistic model BTO, we performed classical molecular dynamics at different temperature for a longer trajectory  of 220 ps and using the standard time step of 1 fs. The different phases of BTO were obtained by  the classical molecular dynamics simulations at different temperatures, namely rhombohedral for T < 80 K, orthorhombic for 85 K < T < 90 K, tetragonal for 95 K < T < 150 K, and cubic  for 155 K < T. These correspond to the experimentally observed phases of BTO, namely  rhombohedral for T < 183 K, orthorhombic for 183 K < T < 278 K, tetragonal for 278 K < T < 393 K, and cubic for 393 K < T. As noted in previous studies, the lower temperatures of the simulated phase transitions compared to the experimental values are due to fundamental inaccuracies of DFT calculations for BTO rather than due to the deficiency of the atomistic model~\cite{BTO_atomistic}.  The  BPM parameters determined  to reproduce the polarizabilities of  10-atom BTO were used to calculate the polarizability trajectories for all phases.  While in this case disagreement between theory and experiment may be due to both the errors in the polarizability calculations as well as the errors due to the atomistic potential, comparison of MD-derived and experimnetal Raman spectra will still reveal the usefulness of  the atomistic MD + BPM approach for  modeling and understanding of experimental Raman spectra.

Figs.~\ref{Fig13} (a), (b), (c) and (d) compare the calculated BPM Raman spectra and experimental spectra  for the rhombohedral, orthorhombic, tetragonal and cubic phases of BTO, respectively, in the  0-700 cm$^{-1}$ range.  Examination of the figures shows that the BVMD-BPM spectra qualitatively  reproduce the experimental peaks and their changes with temperature.  The additional peak above 700 cm$^{-1}$ of the experimental range is also reproduced by BPM for the rhombohedral, orthorhombic and tetragonal phase (see SM). 
However, differences in the peak positions are observed for the BVMD-BPM peaks at lower frequencies in some cases (e.g. peaks 1 and 2 for R-phase, and peak 1 of the C-phase), and at higher frequencies in other cases (e.g. peaks 3-6 for R-phase  and peaks 1-4 for the  O phase).  Considering that BPM reproduces the peak position of DFT spectra, as discussed above,  it is likely that these differences are due to the error introduced by the use of the atomistic potential.  Furthermore, the lineshapes and intensities of the peaks show differences for all phases, with particularly significant differences observed for the C-phase and T-phase peaks 1 and 2, and O-phase peak 2. This is consistent with the above-discussed errors induced in the peak intensity and lineshape  by the errors in the polarizability trajectory.

To further examine the usefulness of the BVMD-BPM approach for BTO simulations, we compared experimental central peak spectra for bulk BTO and a thin BTO film clamped to the substrate (Fig.~\ref{Fig14}) to the BVMD-BPM spectra obtained from MD simulations of bulk BTO and a BTO supercell clamped in the $x$ and $y$ directions (i.e. fixed $x$- and $y$- lattice parameters while the $z$-axis lattice parameter is left free to vary), thus simulating the clamping of the film by the substrate.  The BVMD simulations are performed at 130~K. At this temperature, BVMD simulations obtain the tetragonal phase of BTO corresponding to the experimentally observed tetragonal phase at room temperature.   

Fig.~\ref{Fig14} (left panel) shows that the central peak intensity as calculated from BPM is decreased for the thin-film structure in comparison to bulk structure, in agreement with the central peak intensity change observed experimentally. This indicate that the change in the central peak intensity is due to the clamping of the film by the substrate, rather than by nanoscale effects or domain walls, because neither domain walls nor free surfaces are present in the BVMD simulations.  Taken together with the results presented in Figs.~\ref{Fig13} and ~\ref{Fig14}, these results show that BPM can be used for qualitative modeling of the BTO Raman spectra and their  interpretation. However, fine features of the spectrum  cannot be reproduced with the BPM.
\begin{figure}[]
\centering
\includegraphics[width = 8.5 cm,angle =0]{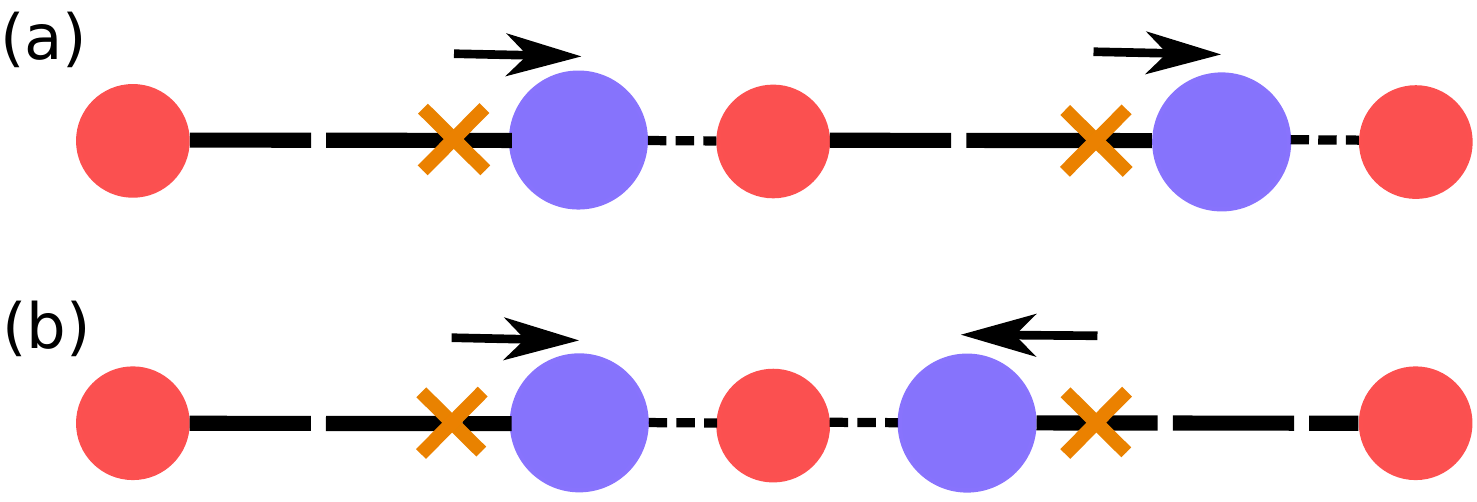}
\caption{(Color online) In a one-dimensional chain of O-Ti-O-Ti-O, (a) head-to-tail, (b) head-to-head, movement of dipole formed due to the displacement of Ti from the center of O as marked by yellow cross. Red and blue circles corresponding to O and Ti, respectively. Each arrow indicate the movement of the dipole.}
\label{Fig16}
\end{figure}
Next, we considered the CPB  perovskite, which due to its soft bonds and large distortions at even relatively low temperature poses a challenge for atomistic polarizability modeling.  Fig.~\ref{Fig15} (a) compares the BPM-calculated polarizability with the DFT-calculated polarizability of CPB at 300 K.  It can be seen that a high scatter that is much stronger than that for BTO  is observed even for the BPM-2$^{nd}$ data.  Similarly, the polarizability trajectories show stronger deviations of the BPM-2$^{nd}$ results from the DFT results compared to BTO.  Comparison of the DFT and BPM-2$^{nd}$ Raman spectra shows that even in this case, BPM-2$^{nd}$ results match the peak position of the DFT data. However, the agreement between the relative intensities is quite poor.  


Consideration of the perovskite structures and the possible distortions shows that there are several sources of error in the BPM for the modeling of BTO, CPB and other perovskites. The most basic shortcoming is that the strong distortions in these systems may require higher-order terms in the Taylor expansion  of bond polarizability to reproduce the DFT results. However, results obtained 3rd- and 4th-order BP models for BTO show only a slight improvement in agreement with DFT results (see SM). Second, similar to the molecular systems, the independent bond approximation of the BPM  ignores the effect of Ti-O-Ti and  O-Ti-O (Pb-Br-Pb and  Br-Pb-Br) angle changes in BTO (CPB). Third, due to its dependence on the  bond lengths only, the BPM cannot distinguish between the head-to-head and head-to-tail dipole structures created by the displacements of the B-cations from the centers of their octahedra, as illustrated in Fig.~\ref{Fig16}. The second and third effects cannot be addressed in the BPM framework due to its fundamental assumption of non-interacting independent contributions of each bond to the total polarizability.   In fact, these effects are  the primary origin of the disagreement between the DFT and BPM polarizability results for BTO and CBP as demonstrated by the weak improvement provided by the BPM-3$^{rd}$ and BPM-4$^{th}$ models for BTO (see SM). 
Thus, the BPM must be fundamentally modified to include the interactions between bonds to achieve accurate prediction of polarizability for solid-state perovskite systems.

\section{CONCLUSIONS}

We have investigated the applicability of the bond polarizability model to the modeling of polarizability and Raman spectra using  trajectories obtained from MD simulations. We find that BPM including second-order terms generally  gives  qualitative agreement with the DFT results such that the positions of the Raman peaks are reproduced accurately but their intensities can differ from those obtained by quantum-mechanical calculations. Structures with higher symmetry and with angular vibrations for which the  polarizability changes  are second-order in the angle change  (e.g. CH$_{4}$)  show much higher accuracy  than highly asymmetric molecules (most importantly water) for which the effect of angle changes on the polarizability is linear in the angle change. Similarly, BPM accuracy is much higher for molecules with stiff angular potential energy surface due to $sp^{2}$ bonding (e.g. CH$_{2}$O and thiophene). For solid-state BTO perovskite, qualitative accuracy is obtained for the BPM-2$^{nd}$ model while for the much softer and distorted CPB,  BPM shows poor agreement.  Due to its simplicity, BPM provides a good starting point for simulations of Raman spectra and can be relied upon to reproduce and interpret strong changes in the experimental Raman spectra. However, the shortcomings of the BPM due to the assumption of non-interacting bonds make it incapable of obtaining quantitative accuracy except for highly symmetric systems and lead to complete failure in case of systems where large deviations from the ground state structure are present. Thus, to enable reliable interpretation of fine features of Raman spectra, it is necessary to develop atomistic models for polarizability that can include the effects of interactions between bonds and angles.

\begin{acknowledgments}
A.P., A.R., S. M., J.E.S. and I.G. acknowledge the support of the Army Research Office
under Grant W911NF-21-1-0126 and Army/ARL via the
Collaborative for Hierarchical Agile and Responsive Materials (CHARM) under cooperative
agreement W911NF-19-2-0119. A.P. and I.G acknowledge additional support from Israel Science Foundation under Grant 1479/21. 
\end{acknowledgments}

\section*{Data Availability}
The data used in this work are available from the corresponding author upon reasonable request.

\bibliographystyle{apsrev4-1}

\begin{thebibliography}{29}%
\makeatletter
\providecommand \@ifxundefined [1]{%
 \@ifx{#1\undefined}
}%
\providecommand \@ifnum [1]{%
 \ifnum #1\expandafter \@firstoftwo
 \else \expandafter \@secondoftwo
 \fi
}%
\providecommand \@ifx [1]{%
 \ifx #1\expandafter \@firstoftwo
 \else \expandafter \@secondoftwo
 \fi
}%
\providecommand \natexlab [1]{#1}%
\providecommand \enquote  [1]{``#1''}%
\providecommand \bibnamefont  [1]{#1}%
\providecommand \bibfnamefont [1]{#1}%
\providecommand \citenamefont [1]{#1}%
\providecommand \href@noop [0]{\@secondoftwo}%
\providecommand \href [0]{\begingroup \@sanitize@url \@href}%
\providecommand \@href[1]{\@@startlink{#1}\@@href}%
\providecommand \@@href[1]{\endgroup#1\@@endlink}%
\providecommand \@sanitize@url [0]{\catcode `\\12\catcode `\$12\catcode
  `\&12\catcode `\#12\catcode `\^12\catcode `\_12\catcode `\%12\relax}%
\providecommand \@@startlink[1]{}%
\providecommand \@@endlink[0]{}%
\providecommand \url  [0]{\begingroup\@sanitize@url \@url }%
\providecommand \@url [1]{\endgroup\@href {#1}{\urlprefix }}%
\providecommand \urlprefix  [0]{URL }%
\providecommand \Eprint [0]{\href }%
\providecommand \doibase [0]{http://dx.doi.org/}%
\providecommand \selectlanguage [0]{\@gobble}%
\providecommand \bibinfo  [0]{\@secondoftwo}%
\providecommand \bibfield  [0]{\@secondoftwo}%
\providecommand \translation [1]{[#1]}%
\providecommand \BibitemOpen [0]{}%
\providecommand \bibitemStop [0]{}%
\providecommand \bibitemNoStop [0]{.\EOS\space}%
\providecommand \EOS [0]{\spacefactor3000\relax}%
\providecommand \BibitemShut  [1]{\csname bibitem#1\endcsname}%
\let\auto@bib@innerbib\@empty
\bibitem [{\citenamefont {Orlando}\ \emph {et~al.}(2021)\citenamefont
  {Orlando}, \citenamefont {Franceschini}, \citenamefont {Muscas},
  \citenamefont {Pidkova}, \citenamefont {Bartoli}, \citenamefont {Rovere},\
  and\ \citenamefont {Tagliaferro}}]{chemosensors9090262}%
  \BibitemOpen
  \bibfield  {author} {\bibinfo {author} {\bibfnamefont {A.}~\bibnamefont
  {Orlando}}, \bibinfo {author} {\bibfnamefont {F.}~\bibnamefont
  {Franceschini}}, \bibinfo {author} {\bibfnamefont {C.}~\bibnamefont
  {Muscas}}, \bibinfo {author} {\bibfnamefont {S.}~\bibnamefont {Pidkova}},
  \bibinfo {author} {\bibfnamefont {M.}~\bibnamefont {Bartoli}}, \bibinfo
  {author} {\bibfnamefont {M.}~\bibnamefont {Rovere}}, \ and\ \bibinfo {author}
  {\bibfnamefont {A.}~\bibnamefont {Tagliaferro}},\ }\href {\doibase
  10.3390/chemosensors9090262} {\bibfield  {journal} {\bibinfo  {journal}
  {Chemosensors}\ }\textbf {\bibinfo {volume} {9}},\ \bibinfo {pages} {262}
  (\bibinfo {year} {2021})}\BibitemShut {NoStop}%
\bibitem [{\citenamefont {Long}(2002)}]{long}%
  \BibitemOpen
  \bibfield  {author} {\bibinfo {author} {\bibfnamefont {D.~A.}\ \bibnamefont
  {Long}},\ }\href {\doibase https://doi.org/10.1002/0470845767.ch1} {\emph
  {\bibinfo {title} {The Raman Effect}}}\ (\bibinfo  {publisher} {John Wiley \&
  Sons, Ltd},\ \bibinfo {year} {2002})\BibitemShut {NoStop}%
\bibitem [{\citenamefont {Gastegger}\ \emph {et~al.}(2017)\citenamefont
  {Gastegger}, \citenamefont {Behler},\ and\ \citenamefont
  {Marquetand}}]{C7SC02267K}%
  \BibitemOpen
  \bibfield  {author} {\bibinfo {author} {\bibfnamefont {M.}~\bibnamefont
  {Gastegger}}, \bibinfo {author} {\bibfnamefont {J.}~\bibnamefont {Behler}}, \
  and\ \bibinfo {author} {\bibfnamefont {P.}~\bibnamefont {Marquetand}},\
  }\href {\doibase 10.1039/C7SC02267K} {\bibfield  {journal} {\bibinfo
  {journal} {Chem. Sci.}\ }\textbf {\bibinfo {volume} {8}},\ \bibinfo {pages}
  {6924} (\bibinfo {year} {2017})}\BibitemShut {NoStop}%
\bibitem [{\citenamefont {Gaigeot}\ and\ \citenamefont {Sprik}(2003)}]{IR_MD}%
  \BibitemOpen
  \bibfield  {author} {\bibinfo {author} {\bibfnamefont {M.-P.}\ \bibnamefont
  {Gaigeot}}\ and\ \bibinfo {author} {\bibfnamefont {M.}~\bibnamefont
  {Sprik}},\ }\href {\doibase 10.1021/jp034788u} {\bibfield  {journal}
  {\bibinfo  {journal} {The Journal of Physical Chemistry B}\ }\textbf
  {\bibinfo {volume} {107}},\ \bibinfo {pages} {10344} (\bibinfo {year}
  {2003})}\BibitemShut {NoStop}%
\bibitem [{\citenamefont {Thomas}\ \emph {et~al.}(2013)\citenamefont {Thomas},
  \citenamefont {Brehm}, \citenamefont {Fligg}, \citenamefont {Vöhringer},\
  and\ \citenamefont {Kirchner}}]{Raman_intensity}%
  \BibitemOpen
  \bibfield  {author} {\bibinfo {author} {\bibfnamefont {M.}~\bibnamefont
  {Thomas}}, \bibinfo {author} {\bibfnamefont {M.}~\bibnamefont {Brehm}},
  \bibinfo {author} {\bibfnamefont {R.}~\bibnamefont {Fligg}}, \bibinfo
  {author} {\bibfnamefont {P.}~\bibnamefont {Vöhringer}}, \ and\ \bibinfo
  {author} {\bibfnamefont {B.}~\bibnamefont {Kirchner}},\ }\href {\doibase
  10.1039/C3CP44302G} {\bibfield  {journal} {\bibinfo  {journal} {Phys. Chem.
  Chem. Phys.}\ }\textbf {\bibinfo {volume} {15}},\ \bibinfo {pages} {6608}
  (\bibinfo {year} {2013})}\BibitemShut {NoStop}%
\bibitem [{\citenamefont {Putrino}\ and\ \citenamefont
  {Parrinello}(2002)}]{PhysRevLett.88.176401}%
  \BibitemOpen
  \bibfield  {author} {\bibinfo {author} {\bibfnamefont {A.}~\bibnamefont
  {Putrino}}\ and\ \bibinfo {author} {\bibfnamefont {M.}~\bibnamefont
  {Parrinello}},\ }\href {\doibase 10.1103/PhysRevLett.88.176401} {\bibfield
  {journal} {\bibinfo  {journal} {Phys. Rev. Lett.}\ }\textbf {\bibinfo
  {volume} {88}},\ \bibinfo {pages} {176401} (\bibinfo {year}
  {2002})}\BibitemShut {NoStop}%
\bibitem [{\citenamefont {Luber}\ \emph {et~al.}(2014)\citenamefont {Luber},
  \citenamefont {Iannuzzi},\ and\ \citenamefont {Hutter}}]{hutter}%
  \BibitemOpen
  \bibfield  {author} {\bibinfo {author} {\bibfnamefont {S.}~\bibnamefont
  {Luber}}, \bibinfo {author} {\bibfnamefont {M.}~\bibnamefont {Iannuzzi}}, \
  and\ \bibinfo {author} {\bibfnamefont {J.}~\bibnamefont {Hutter}},\ }\href
  {\doibase 10.1063/1.4894425} {\bibfield  {journal} {\bibinfo  {journal} {The
  Journal of Chemical Physics}\ }\textbf {\bibinfo {volume} {141}},\ \bibinfo
  {pages} {094503} (\bibinfo {year} {2014})}\BibitemShut {NoStop}%
\bibitem [{\citenamefont {Mattiat}\ and\ \citenamefont
  {Luber}(2021)}]{acs.jctc.0c00755}%
  \BibitemOpen
  \bibfield  {author} {\bibinfo {author} {\bibfnamefont {J.}~\bibnamefont
  {Mattiat}}\ and\ \bibinfo {author} {\bibfnamefont {S.}~\bibnamefont
  {Luber}},\ }\href {\doibase 10.1021/acs.jctc.0c00755} {\bibfield  {journal}
  {\bibinfo  {journal} {Journal of Chemical Theory and Computation}\ }\textbf
  {\bibinfo {volume} {17}},\ \bibinfo {pages} {344} (\bibinfo {year}
  {2021})}\BibitemShut {NoStop}%
\bibitem [{\citenamefont {Lazzeri}\ and\ \citenamefont
  {Mauri}(2003)}]{PhysRevLett.90.036401}%
  \BibitemOpen
  \bibfield  {author} {\bibinfo {author} {\bibfnamefont {M.}~\bibnamefont
  {Lazzeri}}\ and\ \bibinfo {author} {\bibfnamefont {F.}~\bibnamefont
  {Mauri}},\ }\href {\doibase 10.1103/PhysRevLett.90.036401} {\bibfield
  {journal} {\bibinfo  {journal} {Phys. Rev. Lett.}\ }\textbf {\bibinfo
  {volume} {90}},\ \bibinfo {pages} {036401} (\bibinfo {year}
  {2003})}\BibitemShut {NoStop}%
\bibitem [{\citenamefont {Lussier}\ \emph {et~al.}(2020)\citenamefont
  {Lussier}, \citenamefont {Thibault}, \citenamefont {Charron}, \citenamefont
  {Wallace},\ and\ \citenamefont {Masson}}]{LUSSIER2020115796}%
  \BibitemOpen
  \bibfield  {author} {\bibinfo {author} {\bibfnamefont {F.}~\bibnamefont
  {Lussier}}, \bibinfo {author} {\bibfnamefont {V.}~\bibnamefont {Thibault}},
  \bibinfo {author} {\bibfnamefont {B.}~\bibnamefont {Charron}}, \bibinfo
  {author} {\bibfnamefont {G.~Q.}\ \bibnamefont {Wallace}}, \ and\ \bibinfo
  {author} {\bibfnamefont {J.-F.}\ \bibnamefont {Masson}},\ }\href {\doibase
  https://doi.org/10.1016/j.trac.2019.115796} {\bibfield  {journal} {\bibinfo
  {journal} {TrAC Trends in Analytical Chemistry}\ }\textbf {\bibinfo {volume}
  {124}},\ \bibinfo {pages} {115796} (\bibinfo {year} {2020})}\BibitemShut
  {NoStop}%
\bibitem [{\citenamefont {Sommers}\ \emph {et~al.}(2020)\citenamefont
  {Sommers}, \citenamefont {Calegari~Andrade}, \citenamefont {Zhang},
  \citenamefont {Wang},\ and\ \citenamefont {Car}}]{car}%
  \BibitemOpen
  \bibfield  {author} {\bibinfo {author} {\bibfnamefont {G.~M.}\ \bibnamefont
  {Sommers}}, \bibinfo {author} {\bibfnamefont {M.~F.}\ \bibnamefont
  {Calegari~Andrade}}, \bibinfo {author} {\bibfnamefont {L.}~\bibnamefont
  {Zhang}}, \bibinfo {author} {\bibfnamefont {H.}~\bibnamefont {Wang}}, \ and\
  \bibinfo {author} {\bibfnamefont {R.}~\bibnamefont {Car}},\ }\href {\doibase
  10.1039/D0CP01893G} {\bibfield  {journal} {\bibinfo  {journal} {Phys. Chem.
  Chem. Phys.}\ }\textbf {\bibinfo {volume} {22}},\ \bibinfo {pages} {10592}
  (\bibinfo {year} {2020})}\BibitemShut {NoStop}%
\bibitem [{\citenamefont {Berger}\ \emph {et~al.}(2024)\citenamefont {Berger},
  \citenamefont {Niemelä}, \citenamefont {Lampela}, \citenamefont {Juffer},\
  and\ \citenamefont {Komsa}}]{berger2024raman}%
  \BibitemOpen
  \bibfield  {author} {\bibinfo {author} {\bibfnamefont {E.}~\bibnamefont
  {Berger}}, \bibinfo {author} {\bibfnamefont {J.}~\bibnamefont {Niemelä}},
  \bibinfo {author} {\bibfnamefont {O.}~\bibnamefont {Lampela}}, \bibinfo
  {author} {\bibfnamefont {A.~H.}\ \bibnamefont {Juffer}}, \ and\ \bibinfo
  {author} {\bibfnamefont {H.-P.}\ \bibnamefont {Komsa}},\ }\href@noop {}
  {\enquote {\bibinfo {title} {Raman spectra of amino acids and peptides from
  machine learning polarizabilities},}\ } (\bibinfo {year} {2024}),\ \Eprint
  {http://arxiv.org/abs/2401.14808} {arXiv:2401.14808 [physics.comp-ph]}
  \BibitemShut {NoStop}%
\bibitem [{\citenamefont {Raimbault}\ \emph {et~al.}(2019)\citenamefont
  {Raimbault}, \citenamefont {Grisafi}, \citenamefont {Ceriotti},\ and\
  \citenamefont {Rossi}}]{Raimbault_2019}%
  \BibitemOpen
  \bibfield  {author} {\bibinfo {author} {\bibfnamefont {N.}~\bibnamefont
  {Raimbault}}, \bibinfo {author} {\bibfnamefont {A.}~\bibnamefont {Grisafi}},
  \bibinfo {author} {\bibfnamefont {M.}~\bibnamefont {Ceriotti}}, \ and\
  \bibinfo {author} {\bibfnamefont {M.}~\bibnamefont {Rossi}},\ }\href
  {\doibase 10.1088/1367-2630/ab4509} {\bibfield  {journal} {\bibinfo
  {journal} {New Journal of Physics}\ }\textbf {\bibinfo {volume} {21}},\
  \bibinfo {pages} {105001} (\bibinfo {year} {2019})}\BibitemShut {NoStop}%
\bibitem [{\citenamefont {Montero}\ and\ \citenamefont {del Rio}(1976)}]{BPM1}%
  \BibitemOpen
  \bibfield  {author} {\bibinfo {author} {\bibfnamefont {S.}~\bibnamefont
  {Montero}}\ and\ \bibinfo {author} {\bibfnamefont {G.}~\bibnamefont {del
  Rio}},\ }\href {\doibase 10.1080/00268977600100271} {\bibfield  {journal}
  {\bibinfo  {journal} {Molecular Physics}\ }\textbf {\bibinfo {volume} {31}},\
  \bibinfo {pages} {357} (\bibinfo {year} {1976})}\BibitemShut {NoStop}%
\bibitem [{\citenamefont {Cardona}\ \emph {et~al.}()\citenamefont {Cardona},
  \citenamefont {Chang}, \citenamefont {G{\"u}ntherodt}, \citenamefont {Long},\
  and\ \citenamefont {Vogt}}]{cardonalight}%
  \BibitemOpen
  \bibfield  {author} {\bibinfo {author} {\bibfnamefont {M.}~\bibnamefont
  {Cardona}}, \bibinfo {author} {\bibfnamefont {R.}~\bibnamefont {Chang}},
  \bibinfo {author} {\bibfnamefont {G.}~\bibnamefont {G{\"u}ntherodt}},
  \bibinfo {author} {\bibfnamefont {M.}~\bibnamefont {Long}}, \ and\ \bibinfo
  {author} {\bibfnamefont {H.}~\bibnamefont {Vogt}},\ }\href
  {https://doi.org/10.1007/3-540-11380-0} {\emph {\bibinfo {title} {Light
  Scattering in Solids II: Basic Concepts and Instrumentation}}},\ Topics in
  Applied Physics\ (\bibinfo  {publisher} {Springer-Verlag Berlin Heidelbergi
  1982})\BibitemShut {NoStop}%
\bibitem [{199(1996)}]{1996273}%
  \BibitemOpen
  in\ \href {\doibase https://doi.org/10.1016/S0090-1911(96)80014-X} {\emph
  {\bibinfo {booktitle} {Vibrational Intensities}}},\ \bibinfo {series}
  {Vibrational Spectra and Structure}, Vol.~\bibinfo {volume} {22},\ \bibinfo
  {editor} {edited by\ \bibinfo {editor} {\bibfnamefont {B.~S.}\ \bibnamefont
  {Galabov}}\ and\ \bibinfo {editor} {\bibfnamefont {T.}~\bibnamefont
  {Dudev}}}\ (\bibinfo  {publisher} {Elsevier},\ \bibinfo {year} {1996})\ pp.\
  \bibinfo {pages} {215--271}\BibitemShut {NoStop}%
\bibitem [{\citenamefont {Liang}\ \emph {et~al.}(2017)\citenamefont {Liang},
  \citenamefont {Puretzky}, \citenamefont {Sumpter},\ and\ \citenamefont
  {Meunier}}]{C7NR05839J}%
  \BibitemOpen
  \bibfield  {author} {\bibinfo {author} {\bibfnamefont {L.}~\bibnamefont
  {Liang}}, \bibinfo {author} {\bibfnamefont {A.~A.}\ \bibnamefont {Puretzky}},
  \bibinfo {author} {\bibfnamefont {B.~G.}\ \bibnamefont {Sumpter}}, \ and\
  \bibinfo {author} {\bibfnamefont {V.}~\bibnamefont {Meunier}},\ }\href
  {\doibase 10.1039/C7NR05839J} {\bibfield  {journal} {\bibinfo  {journal}
  {Nanoscale}\ }\textbf {\bibinfo {volume} {9}},\ \bibinfo {pages} {15340}
  (\bibinfo {year} {2017})}\BibitemShut {NoStop}%
\bibitem [{\citenamefont {Wirtz}\ \emph {et~al.}(2005)\citenamefont {Wirtz},
  \citenamefont {Lazzeri}, \citenamefont {Mauri},\ and\ \citenamefont
  {Rubio}}]{PhysRevB.71.241402}%
  \BibitemOpen
  \bibfield  {author} {\bibinfo {author} {\bibfnamefont {L.}~\bibnamefont
  {Wirtz}}, \bibinfo {author} {\bibfnamefont {M.}~\bibnamefont {Lazzeri}},
  \bibinfo {author} {\bibfnamefont {F.}~\bibnamefont {Mauri}}, \ and\ \bibinfo
  {author} {\bibfnamefont {A.}~\bibnamefont {Rubio}},\ }\href {\doibase
  10.1103/PhysRevB.71.241402} {\bibfield  {journal} {\bibinfo  {journal} {Phys.
  Rev. B}\ }\textbf {\bibinfo {volume} {71}},\ \bibinfo {pages} {241402}
  (\bibinfo {year} {2005})}\BibitemShut {NoStop}%
\bibitem [{\citenamefont {Luo}\ \emph {et~al.}(2015)\citenamefont {Luo},
  \citenamefont {Lu}, \citenamefont {Cong}, \citenamefont {Yu}, \citenamefont
  {Xiong},\ and\ \citenamefont {Ying~Quek}}]{Luo2015}%
  \BibitemOpen
  \bibfield  {author} {\bibinfo {author} {\bibfnamefont {X.}~\bibnamefont
  {Luo}}, \bibinfo {author} {\bibfnamefont {X.}~\bibnamefont {Lu}}, \bibinfo
  {author} {\bibfnamefont {C.}~\bibnamefont {Cong}}, \bibinfo {author}
  {\bibfnamefont {T.}~\bibnamefont {Yu}}, \bibinfo {author} {\bibfnamefont
  {Q.}~\bibnamefont {Xiong}}, \ and\ \bibinfo {author} {\bibfnamefont
  {S.}~\bibnamefont {Ying~Quek}},\ }\href {\doibase 10.1038/srep14565}
  {\bibfield  {journal} {\bibinfo  {journal} {Scientific Reports}\ }\textbf
  {\bibinfo {volume} {5}},\ \bibinfo {pages} {14565} (\bibinfo {year}
  {2015})}\BibitemShut {NoStop}%
\bibitem [{\citenamefont {Umari}\ \emph {et~al.}(2001)\citenamefont {Umari},
  \citenamefont {Pasquarello},\ and\ \citenamefont
  {Dal~Corso}}]{PhysRevB.63.094305}%
  \BibitemOpen
  \bibfield  {author} {\bibinfo {author} {\bibfnamefont {P.}~\bibnamefont
  {Umari}}, \bibinfo {author} {\bibfnamefont {A.}~\bibnamefont {Pasquarello}},
  \ and\ \bibinfo {author} {\bibfnamefont {A.}~\bibnamefont {Dal~Corso}},\
  }\href {\doibase 10.1103/PhysRevB.63.094305} {\bibfield  {journal} {\bibinfo
  {journal} {Phys. Rev. B}\ }\textbf {\bibinfo {volume} {63}},\ \bibinfo
  {pages} {094305} (\bibinfo {year} {2001})}\BibitemShut {NoStop}%
\bibitem [{\citenamefont {Hermet}\ \emph {et~al.}(2006)\citenamefont {Hermet},
  \citenamefont {Izard}, \citenamefont {Rahmani},\ and\ \citenamefont
  {Ghosez}}]{ghosez}%
  \BibitemOpen
  \bibfield  {author} {\bibinfo {author} {\bibfnamefont {P.}~\bibnamefont
  {Hermet}}, \bibinfo {author} {\bibfnamefont {N.}~\bibnamefont {Izard}},
  \bibinfo {author} {\bibfnamefont {A.}~\bibnamefont {Rahmani}}, \ and\
  \bibinfo {author} {\bibfnamefont {P.}~\bibnamefont {Ghosez}},\ }\href
  {\doibase 10.1021/jp064700n} {\bibfield  {journal} {\bibinfo  {journal} {The
  Journal of Physical Chemistry B}\ }\textbf {\bibinfo {volume} {110}},\
  \bibinfo {pages} {24869} (\bibinfo {year} {2006})}\BibitemShut {NoStop}%
\bibitem [{\citenamefont {Berger}\ and\ \citenamefont
  {Komsa}(2023)}]{berger2023polarizability}%
  \BibitemOpen
  \bibfield  {author} {\bibinfo {author} {\bibfnamefont {E.}~\bibnamefont
  {Berger}}\ and\ \bibinfo {author} {\bibfnamefont {H.-P.}\ \bibnamefont
  {Komsa}},\ }\href@noop {} {\enquote {\bibinfo {title} {Polarizability models
  for simulations of finite temperature raman spectra from machine learning
  molecular dynamics},}\ } (\bibinfo {year} {2023}),\ \Eprint
  {http://arxiv.org/abs/2310.13310} {arXiv:2310.13310 [cond-mat.mes-hall]}
  \BibitemShut {NoStop}%
\bibitem [{\citenamefont {Baroni}\ \emph {et~al.}(2001)\citenamefont {Baroni},
  \citenamefont {de~Gironcoli}, \citenamefont {Dal~Corso},\ and\ \citenamefont
  {Giannozzi}}]{DFPT_baroni}%
  \BibitemOpen
  \bibfield  {author} {\bibinfo {author} {\bibfnamefont {S.}~\bibnamefont
  {Baroni}}, \bibinfo {author} {\bibfnamefont {S.}~\bibnamefont
  {de~Gironcoli}}, \bibinfo {author} {\bibfnamefont {A.}~\bibnamefont
  {Dal~Corso}}, \ and\ \bibinfo {author} {\bibfnamefont {P.}~\bibnamefont
  {Giannozzi}},\ }\href {\doibase 10.1103/RevModPhys.73.515} {\bibfield
  {journal} {\bibinfo  {journal} {Rev. Mod. Phys.}\ }\textbf {\bibinfo {volume}
  {73}},\ \bibinfo {pages} {515} (\bibinfo {year} {2001})}\BibitemShut
  {NoStop}%
\bibitem [{\citenamefont {Giannozzi}\ \emph {et~al.}(2009)\citenamefont
  {Giannozzi}, \citenamefont {Baroni}, \citenamefont {Bonini}, \citenamefont
  {Calandra}, \citenamefont {Car}, \citenamefont {Cavazzoni}, \citenamefont
  {Ceresoli}, \citenamefont {Chiarotti}, \citenamefont {Cococcioni},
  \citenamefont {Dabo}, \citenamefont {Corso}, \citenamefont {de~Gironcoli},
  \citenamefont {Fabris}, \citenamefont {Fratesi}, \citenamefont {Gebauer},
  \citenamefont {Gerstmann}, \citenamefont {Gougoussis}, \citenamefont
  {Kokalj}, \citenamefont {Lazzeri}, \citenamefont {Martin-Samos},
  \citenamefont {Marzari}, \citenamefont {Mauri}, \citenamefont {Mazzarello},
  \citenamefont {Paolini}, \citenamefont {Pasquarello}, \citenamefont
  {Paulatto}, \citenamefont {Sbraccia}, \citenamefont {Scandolo}, \citenamefont
  {Sclauzero}, \citenamefont {Seitsonen}, \citenamefont {Smogunov},
  \citenamefont {Umari},\ and\ \citenamefont {Wentzcovitch}}]{QE}%
  \BibitemOpen
  \bibfield  {author} {\bibinfo {author} {\bibfnamefont {P.}~\bibnamefont
  {Giannozzi}}, \bibinfo {author} {\bibfnamefont {S.}~\bibnamefont {Baroni}},
  \bibinfo {author} {\bibfnamefont {N.}~\bibnamefont {Bonini}}, \bibinfo
  {author} {\bibfnamefont {M.}~\bibnamefont {Calandra}}, \bibinfo {author}
  {\bibfnamefont {R.}~\bibnamefont {Car}}, \bibinfo {author} {\bibfnamefont
  {C.}~\bibnamefont {Cavazzoni}}, \bibinfo {author} {\bibfnamefont
  {D.}~\bibnamefont {Ceresoli}}, \bibinfo {author} {\bibfnamefont {G.~L.}\
  \bibnamefont {Chiarotti}}, \bibinfo {author} {\bibfnamefont {M.}~\bibnamefont
  {Cococcioni}}, \bibinfo {author} {\bibfnamefont {I.}~\bibnamefont {Dabo}},
  \bibinfo {author} {\bibfnamefont {A.~D.}\ \bibnamefont {Corso}}, \bibinfo
  {author} {\bibfnamefont {S.}~\bibnamefont {de~Gironcoli}}, \bibinfo {author}
  {\bibfnamefont {S.}~\bibnamefont {Fabris}}, \bibinfo {author} {\bibfnamefont
  {G.}~\bibnamefont {Fratesi}}, \bibinfo {author} {\bibfnamefont
  {R.}~\bibnamefont {Gebauer}}, \bibinfo {author} {\bibfnamefont
  {U.}~\bibnamefont {Gerstmann}}, \bibinfo {author} {\bibfnamefont
  {C.}~\bibnamefont {Gougoussis}}, \bibinfo {author} {\bibfnamefont
  {A.}~\bibnamefont {Kokalj}}, \bibinfo {author} {\bibfnamefont
  {M.}~\bibnamefont {Lazzeri}}, \bibinfo {author} {\bibfnamefont
  {L.}~\bibnamefont {Martin-Samos}}, \bibinfo {author} {\bibfnamefont
  {N.}~\bibnamefont {Marzari}}, \bibinfo {author} {\bibfnamefont
  {F.}~\bibnamefont {Mauri}}, \bibinfo {author} {\bibfnamefont
  {R.}~\bibnamefont {Mazzarello}}, \bibinfo {author} {\bibfnamefont
  {S.}~\bibnamefont {Paolini}}, \bibinfo {author} {\bibfnamefont
  {A.}~\bibnamefont {Pasquarello}}, \bibinfo {author} {\bibfnamefont
  {L.}~\bibnamefont {Paulatto}}, \bibinfo {author} {\bibfnamefont
  {C.}~\bibnamefont {Sbraccia}}, \bibinfo {author} {\bibfnamefont
  {S.}~\bibnamefont {Scandolo}}, \bibinfo {author} {\bibfnamefont
  {G.}~\bibnamefont {Sclauzero}}, \bibinfo {author} {\bibfnamefont {A.~P.}\
  \bibnamefont {Seitsonen}}, \bibinfo {author} {\bibfnamefont {A.}~\bibnamefont
  {Smogunov}}, \bibinfo {author} {\bibfnamefont {P.}~\bibnamefont {Umari}}, \
  and\ \bibinfo {author} {\bibfnamefont {R.~M.}\ \bibnamefont {Wentzcovitch}},\
  }\href {\doibase 10.1088/0953-8984/21/39/395502} {\bibfield  {journal}
  {\bibinfo  {journal} {Journal of Physics: Condensed Matter}\ }\textbf
  {\bibinfo {volume} {21}},\ \bibinfo {pages} {395502} (\bibinfo {year}
  {2009})}\BibitemShut {NoStop}%
\bibitem [{\citenamefont {Perdew}\ \emph {et~al.}(2008)\citenamefont {Perdew},
  \citenamefont {Ruzsinszky}, \citenamefont {Csonka}, \citenamefont {Vydrov},
  \citenamefont {Scuseria}, \citenamefont {Constantin}, \citenamefont {Zhou},\
  and\ \citenamefont {Burke}}]{PBEsol}%
  \BibitemOpen
  \bibfield  {author} {\bibinfo {author} {\bibfnamefont {J.~P.}\ \bibnamefont
  {Perdew}}, \bibinfo {author} {\bibfnamefont {A.}~\bibnamefont {Ruzsinszky}},
  \bibinfo {author} {\bibfnamefont {G.~I.}\ \bibnamefont {Csonka}}, \bibinfo
  {author} {\bibfnamefont {O.~A.}\ \bibnamefont {Vydrov}}, \bibinfo {author}
  {\bibfnamefont {G.~E.}\ \bibnamefont {Scuseria}}, \bibinfo {author}
  {\bibfnamefont {L.~A.}\ \bibnamefont {Constantin}}, \bibinfo {author}
  {\bibfnamefont {X.}~\bibnamefont {Zhou}}, \ and\ \bibinfo {author}
  {\bibfnamefont {K.}~\bibnamefont {Burke}},\ }\href {\doibase
  10.1103/PhysRevLett.100.136406} {\bibfield  {journal} {\bibinfo  {journal}
  {Phys. Rev. Lett.}\ }\textbf {\bibinfo {volume} {100}},\ \bibinfo {pages}
  {136406} (\bibinfo {year} {2008})}\BibitemShut {NoStop}%
\bibitem [{\citenamefont {Thompson}\ \emph {et~al.}(2022)\citenamefont
  {Thompson}, \citenamefont {Aktulga}, \citenamefont {Berger}, \citenamefont
  {Bolintineanu}, \citenamefont {Brown}, \citenamefont {Crozier}, \citenamefont
  {in~'t Veld}, \citenamefont {Kohlmeyer}, \citenamefont {Moore}, \citenamefont
  {Nguyen}, \citenamefont {Shan}, \citenamefont {Stevens}, \citenamefont
  {Tranchida}, \citenamefont {Trott},\ and\ \citenamefont {Plimpton}}]{LAMMPS}%
  \BibitemOpen
  \bibfield  {author} {\bibinfo {author} {\bibfnamefont {A.~P.}\ \bibnamefont
  {Thompson}}, \bibinfo {author} {\bibfnamefont {H.~M.}\ \bibnamefont
  {Aktulga}}, \bibinfo {author} {\bibfnamefont {R.}~\bibnamefont {Berger}},
  \bibinfo {author} {\bibfnamefont {D.~S.}\ \bibnamefont {Bolintineanu}},
  \bibinfo {author} {\bibfnamefont {W.~M.}\ \bibnamefont {Brown}}, \bibinfo
  {author} {\bibfnamefont {P.~S.}\ \bibnamefont {Crozier}}, \bibinfo {author}
  {\bibfnamefont {P.~J.}\ \bibnamefont {in~'t Veld}}, \bibinfo {author}
  {\bibfnamefont {A.}~\bibnamefont {Kohlmeyer}}, \bibinfo {author}
  {\bibfnamefont {S.~G.}\ \bibnamefont {Moore}}, \bibinfo {author}
  {\bibfnamefont {T.~D.}\ \bibnamefont {Nguyen}}, \bibinfo {author}
  {\bibfnamefont {R.}~\bibnamefont {Shan}}, \bibinfo {author} {\bibfnamefont
  {M.~J.}\ \bibnamefont {Stevens}}, \bibinfo {author} {\bibfnamefont
  {J.}~\bibnamefont {Tranchida}}, \bibinfo {author} {\bibfnamefont
  {C.}~\bibnamefont {Trott}}, \ and\ \bibinfo {author} {\bibfnamefont {S.~J.}\
  \bibnamefont {Plimpton}},\ }\href {\doibase 10.1016/j.cpc.2021.108171}
  {\bibfield  {journal} {\bibinfo  {journal} {Comp. Phys. Comm.}\ }\textbf
  {\bibinfo {volume} {271}},\ \bibinfo {pages} {108171} (\bibinfo {year}
  {2022})}\BibitemShut {NoStop}%
\bibitem [{\citenamefont {Qi}\ \emph {et~al.}(2016)\citenamefont {Qi},
  \citenamefont {Liu}, \citenamefont {Grinberg},\ and\ \citenamefont
  {Rappe}}]{BTO_atomistic}%
  \BibitemOpen
  \bibfield  {author} {\bibinfo {author} {\bibfnamefont {Y.}~\bibnamefont
  {Qi}}, \bibinfo {author} {\bibfnamefont {S.}~\bibnamefont {Liu}}, \bibinfo
  {author} {\bibfnamefont {I.}~\bibnamefont {Grinberg}}, \ and\ \bibinfo
  {author} {\bibfnamefont {A.~M.}\ \bibnamefont {Rappe}},\ }\href {\doibase
  10.1103/PhysRevB.94.134308} {\bibfield  {journal} {\bibinfo  {journal} {Phys.
  Rev. B}\ }\textbf {\bibinfo {volume} {94}},\ \bibinfo {pages} {134308}
  (\bibinfo {year} {2016})}\BibitemShut {NoStop}%
\bibitem [{\citenamefont {Deluca}\ \emph {et~al.}(2018)\citenamefont {Deluca},
  \citenamefont {Al-Jlaihawi}, \citenamefont {Reichmann}, \citenamefont
  {Bell},\ and\ \citenamefont {Feteira}}]{C7TA11096K}%
  \BibitemOpen
  \bibfield  {author} {\bibinfo {author} {\bibfnamefont {M.}~\bibnamefont
  {Deluca}}, \bibinfo {author} {\bibfnamefont {Z.~G.}\ \bibnamefont
  {Al-Jlaihawi}}, \bibinfo {author} {\bibfnamefont {K.}~\bibnamefont
  {Reichmann}}, \bibinfo {author} {\bibfnamefont {A.~M.~T.}\ \bibnamefont
  {Bell}}, \ and\ \bibinfo {author} {\bibfnamefont {A.}~\bibnamefont
  {Feteira}},\ }\href {\doibase 10.1039/C7TA11096K} {\bibfield  {journal}
  {\bibinfo  {journal} {J. Mater. Chem. A}\ }\textbf {\bibinfo {volume} {6}},\
  \bibinfo {pages} {5443} (\bibinfo {year} {2018})}\BibitemShut {NoStop}%
\bibitem [{\citenamefont {Yuzyuk}(2012)}]{Yuzyuk2012}%
  \BibitemOpen
  \bibfield  {author} {\bibinfo {author} {\bibfnamefont {Y.~I.}\ \bibnamefont
  {Yuzyuk}},\ }\href {\doibase 10.1134/S1063783412050502} {\bibfield  {journal}
  {\bibinfo  {journal} {Physics of the Solid State}\ }\textbf {\bibinfo
  {volume} {54}},\ \bibinfo {pages} {1026} (\bibinfo {year}
  {2012})}\BibitemShut {NoStop}%
\end{thebibliography}
\providecommand{\noopsort}[1]{}\providecommand{\singleletter}[1]{#1}%
\end{document}


\preprint{AIP/123-QED}

\title{Supplementary Material for Accuracy and limitations of the bond polarizability model in modeling of Raman spectra from molecular dynamics simulations}
\author{Atanu Paul}
 \affiliation{Department of Chemistry, Bar-Ilan University, Ramat Gan 5290002, Israel}
\author{Maya Rubenstein}%
\affiliation{Department of Chemistry, Bar-Ilan University, Ramat Gan 5290002, Israel}
x\author{Anthony Ruffino}
\affiliation{Department of Physics, Drexel University, Philadelphia, Pennsylvania 19104, USA}
\author{Stefan Masiuk}
\affiliation{Department of Materials Science and Engineering, Drexel University, Philadelphia, Pennsylvania 19104, USA}
\author{Jonathan Spanier}
\affiliation{Department of Materials Science and Engineering, Drexel University, Philadelphia, Pennsylvania 19104, USA}
\affiliation{Department of Physics, Drexel University, Philadelphia, Pennsylvania 19104, USA}

\author{Ilya Grinberg}
\email[]{ilya.grinberg@biu.ac.il}
\affiliation{%
Department of Chemistry, Bar-Ilan University, Ramat Gan 5290002, Israel
}%


\begin{abstract}

\end{abstract}

\maketitle

\section{Effect of BaO$_{12}$ on polarizability of BaTiO$_{3}$}
\begin{figure}[ht!]
\centering
\includegraphics[width = 8.3 cm,angle =0]{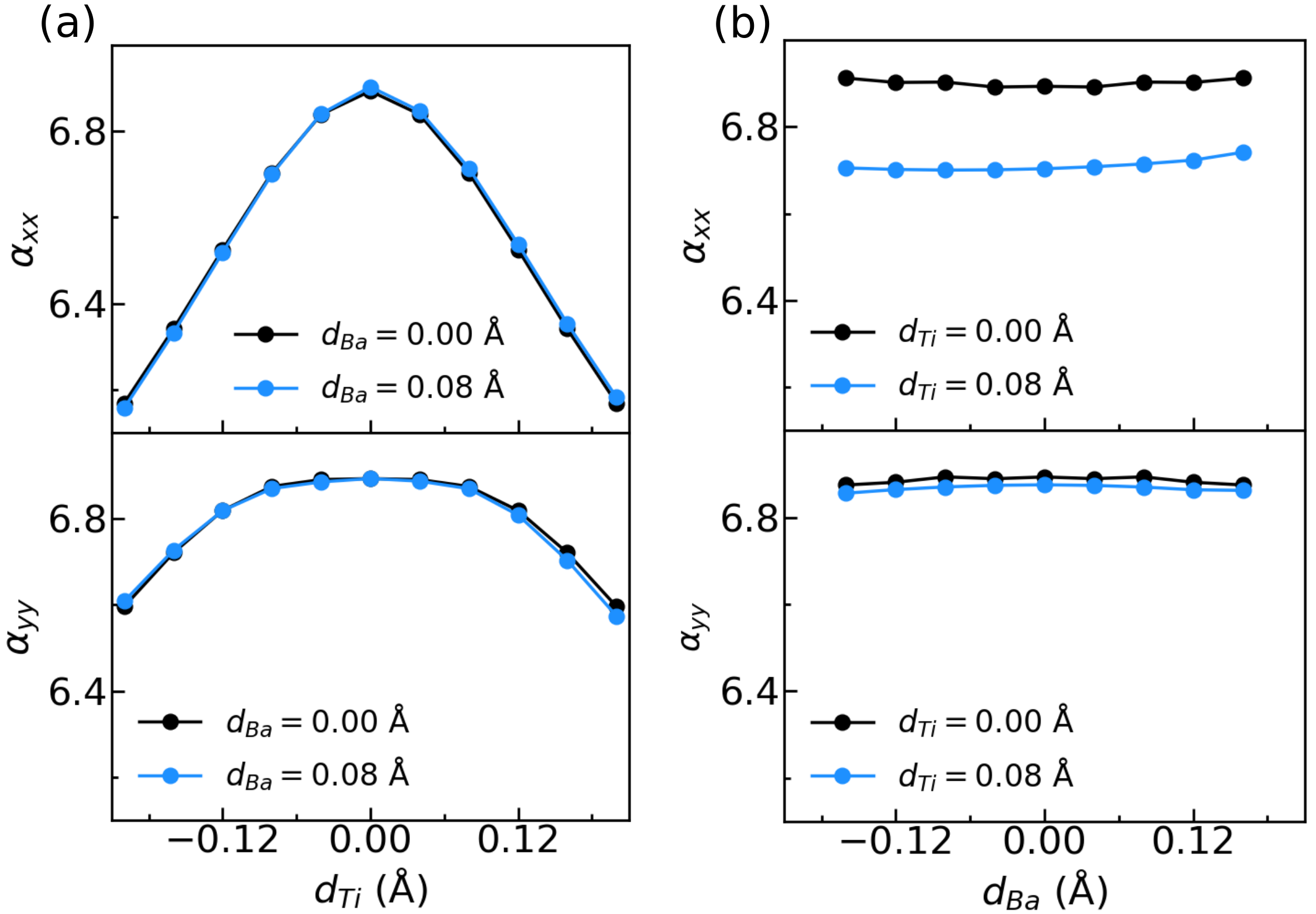}
\caption{(Color online) $\alpha_{xx}$ and $\alpha_{yy}$ of BTO (5-atom cell) calculated from DFT, where $d_{Ba}$ and $d_{Ti}$ are the displacements of Ba and Ti atoms,
respectively, along the [100] direction from their  high-symmetry positions denoted by $d_{Ba}$ = 0 and $d_{Ti}$ = 0. (a) $\alpha_{xx}$ (top panel)
and $\alpha_{xx}$ (bottom panel) as a function of $d_{Ti}$ for two values of
$d_{Ba}$ (0.00 and 0.08 \AA). (b) $\alpha_{xx}$ (top panel) and $\alpha_{yy}$ (bottom
panel) as a function of $d_{Ba}$ for two values of $d_{Ti}$ (0.00 and 0.08 \AA).}
\label{SFig1}
\end{figure}

In order to understand the effect of BaO$_{12}$ on the polarizability of BaTiO$_{3}$ (BTO), we have considered a cubic 5-atom unit cell of BTO and calculated polarizability from DFT for different displacements of Ba and Ti atoms along the [100] direction from their high-symmetry positions. Fig.~S\ref{SFig1} (a) shows the DFT-calculated  $\alpha_{xx}$ and $\alpha_{yy}$ as a function of Ti displacement ($d_{Ti}$, $d_{Ti}$ is the displacement of the Ti atom from its centrosymmetric position) for two Ba positions ($d_{Ba}$ = 0.00 and 0.08 \AA, where $d_{Ba}$ is the displacement of the Ba atom from its centrosymmetric position). The results show a nearly parabolic dependence of $\alpha_{xx}$ and $\alpha_{yy}$ for the displacement of Ti. By contrast, the results for the two different Ba positions ($d_{Ba}$ = 0.00 and 0.08 \AA) show nearly identical $\alpha_{xx}$ and $\alpha_{yy}$ values. This implies that the polarizability of BTO is mostly controlled by the Ti and O movement, and the movement of Ba has only a weak effect  on the  polarizability. In Fig.~S\ref{SFig1} (b), we show the dependence of $\alpha_{xx}$ and $\alpha_{yy}$ as a function of Ba displacement for two different Ti positions  ($d_{Ti}$ = 0.00 and 0.08 \AA). As expected, only weak changes  in the polarizability values are obtained  for Ba displacements. The above observations are due to the fact that Ti atoms are strongly covalently bonded with the O atoms, whereas the bonds between O and Ba atoms are mostly ionic. Therefore, for simplicity, we have neglected the effect of BaO$_{12}$ bonds in our bond polarizability model for BTO.              
\begin{figure}[ht!]
\centering
\includegraphics[width = 8.3 cm,angle =0]{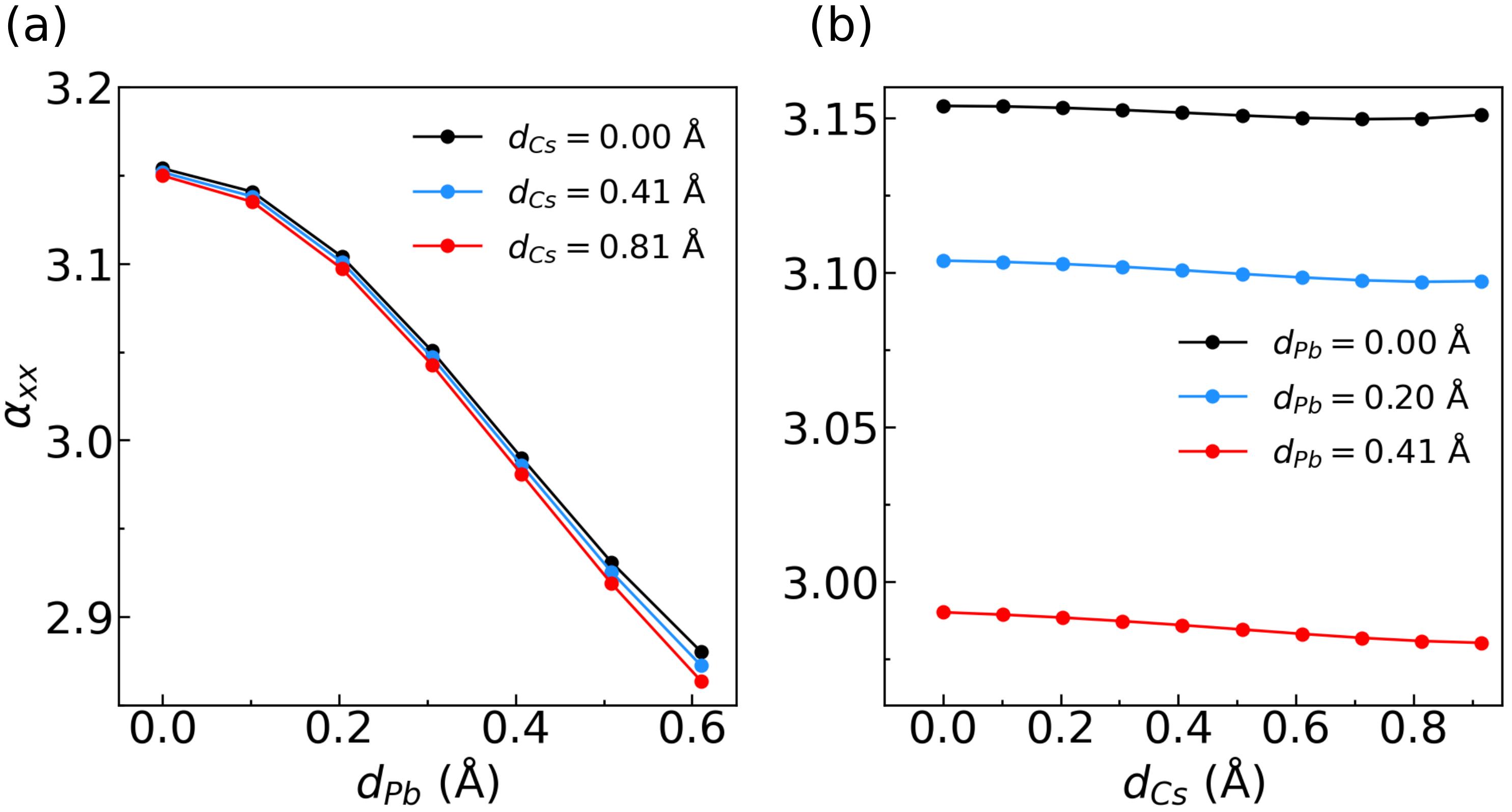}
\caption{(Color online) $\alpha_{xx}$ of CPB (5-atom cell) calculated from DFT, where $d_{Cs}$ and $d_{Pb}$ are the displacements of Cs and Pb atoms,
respectively, along the [111] direction from their high-symmetry positions denoted by $d_{Cs}$ = 0 and $d_{Pb}$ = 0. (a) $\alpha_{xx}$ as a function of $d_{Pb}$ for three values of
$d_{Cs}$ (0.00, 0.41 and 0.81 \AA). (b) $\alpha_{xx}$ as a function of $d_{Cs}$ for three values of $d_{Pb}$ (0.00, 0.20 and 0.41 \AA).}
\label{SFig2}
\end{figure}

\begin{figure*}[ht!]
\centering
\includegraphics[width = 17.5 cm,angle =0]{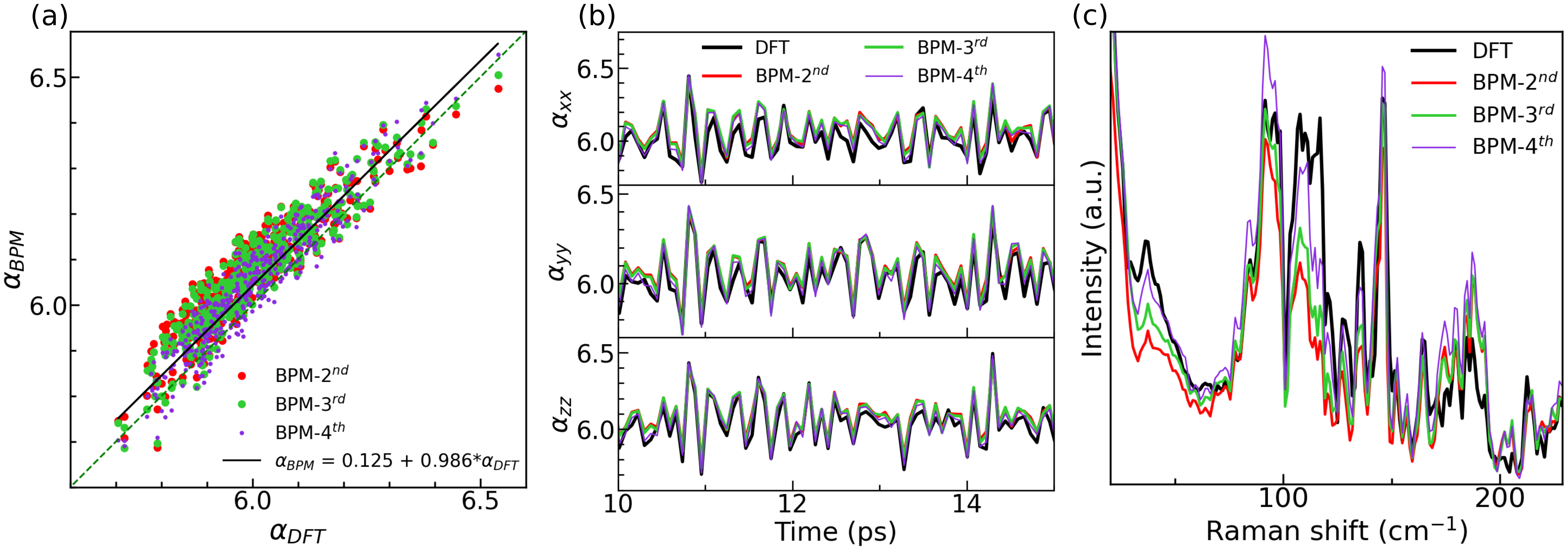}
\caption{(Color online) BaTiO$_{3}$ at 300 K: (Color online) (a) The results for the $xx$ component of $\alpha_{BPM}$ (2$^{nd}$, 3$^{rd}$ and 4$^{th}$ order BPM are shown by
red, green and violet circles, respectively) plotted vs $\alpha_{DFT}$. The linear fit of BPM-4$^{th}$ order data is shown by the black line. The green dotted line
represents $\alpha_{BPM}$ = $\alpha_{DFT}$. (b) Comparison of the trajectory of $\alpha_{xx}$ (top panel), $\alpha_{yy}$ (middle panel) and $\alpha_{zz}$ (bottom panel) calculated from DFT (black), BPM-2$^{nd}$ (red), BPM-3$^{rd}$ (green) and BPM-4$^{th}$ (violet) order. (c) Raman spectra generated using DFT (black), BPM-2$^{nd}$ (red), BPM-3$^{rd}$ (green) and BPM-4$^{th}$ (violet).
}
\label{SFig3}
\end{figure*}

\begin{figure*}[ht!]
\centering
\includegraphics[width = 17.5 cm,angle =0]{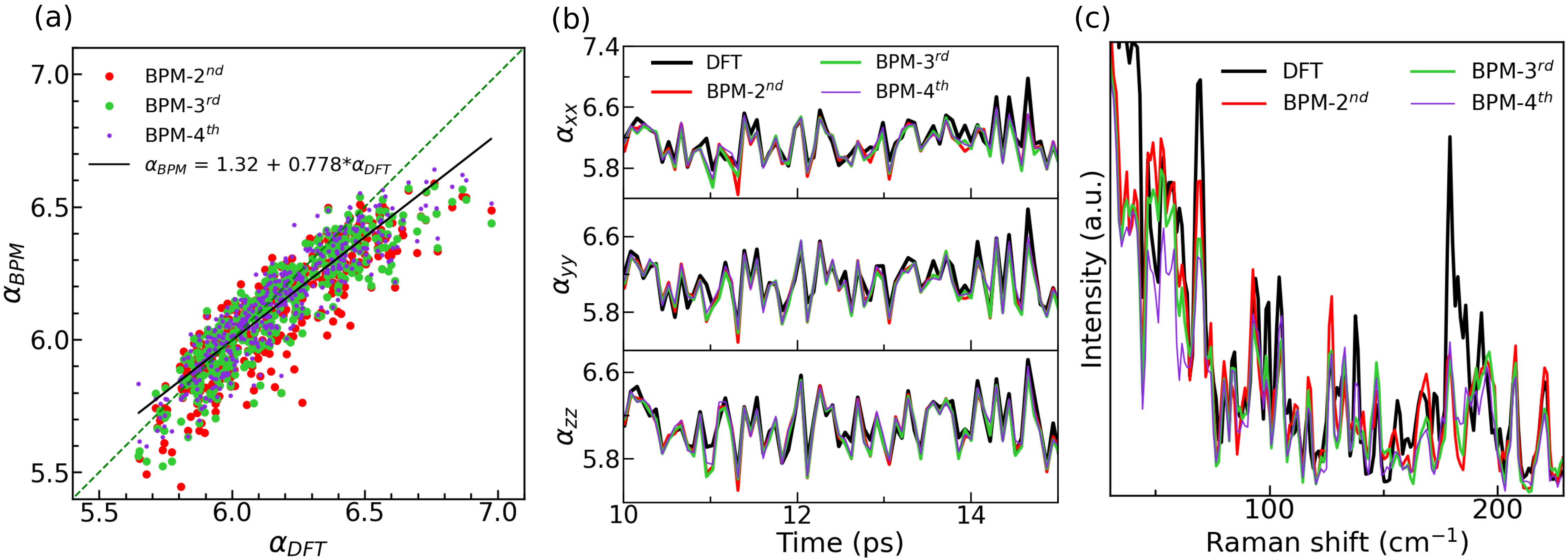}
\caption{(Color online) BaTiO$_{3}$ at 1000 K: (Color online) (a) The results for the $xx$ component of $\alpha_{BPM}$ (2$^{nd}$, 3$^{rd}$ and 4$^{th}$ order BPM are shown by
red, green and violet circles, respectively) plotted vs $\alpha_{DFT}$. The linear fit of BPM-4$^{th}$ order data is shown by the black line. The green dotted line
represents $\alpha_{BPM}$ = $\alpha_{DFT}$. (b) Comparison of the trajectory of $\alpha_{xx}$ (top panel), $\alpha_{yy}$ (middle panel) and $\alpha_{zz}$ (bottom panel) calculated from DFT (black), BPM-2$^{nd}$ (red), BPM-3$^{rd}$ (green) and BPM-4$^{th}$ (violet) order. (c) Raman spectra generated using DFT (black), BPM-2$^{nd}$ (red), BPM-3$^{rd}$ (green) and BPM-4$^{th}$ (violet).}
\label{SFig4}
\end{figure*}
\section{Effect of CsBr$_{12}$ on polarizability of CsPbBr$_{3}$}
We followed the procedure described above for  BTO in order to understand the effect of CsBr$_{12}$ complex on the polarizability of CsPbBr$_{3}$ (CPB). We considered 5-atom cell of CPB and displaced Cs and Pb atoms along the [111] direction from their high-symmetry positions. In Fig.~S\ref{SFig2} (a), we show the dependence of $\alpha_{xx}$ on Pb-displacement from its centrosymmetric position denoted by $d_{Pb}$ for three different Cs positions ($d_{Cs}$ = 0.00, 0.41 and 0.81 \AA, where $d_{Cs}$ be the displacement of the Cs atom from its centrosymmetric position). Similar to BTO, $\alpha_{xx}$ shows a strong dependence on the value of $d_{Pb}$ irrespective of the value of $d_{Cs}$. This indicates that  PbBr$_{6}$ strongly controls the value of the polarizability in CPB with only weak effect of  CsBr$_{12}$ on the polarizability. This is further seen in the Fig.~S\ref{SFig2} (b), where  $\alpha_{xx}$ hardly shows any changes with the Cs displacement ($d_{Cs}$) for the three Pb positions  ($d_{Pb}$ = 0.00, 0.20 and 0.41 \AA). Therefore, we neglected the effect of Cs in the calculation of the polarizability of CPB using BPM.      

\begin{figure*}[ht!]
\centering
\includegraphics[width = 17.5 cm,angle =0]{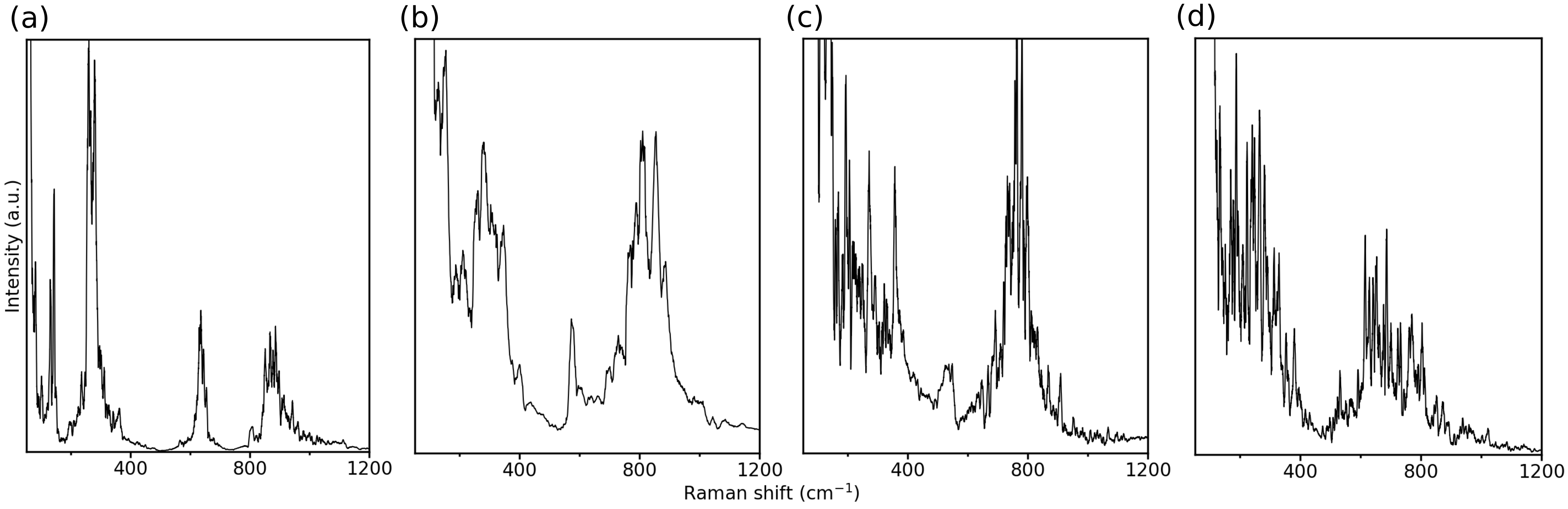}
\caption{(Color online) Raman spectra of different phases ((a) rhombohedral, (b) orthorhombic), (c) tetragonal, and (d) cubic) of BTO single crystal from bond polarizability model (2$^{nd}$ order). The temperatures for (a), (b), (c) and (d) are 60 K, 88 K, 110 K and 160 K, respectively.}
\label{SFig5}
\end{figure*}

\section{BPM-3$^{rd}$ and BPM-4$^{th}$ of BaTiO$_{3}$ at 300 K and 1000 K}
Fig.~S\ref{SFig3} (a) presents the plots of $\alpha_{xx}$ as calculated using BPM-3$^{rd}$ and BPM-4$^{th}$ versus the DFT-calculated $\alpha_{xx}$ for 10-atom BTO along the polarizability trajectory at 300 K. For better understanding of the effect of the inclusion of the higher-order-BPM, we have also shown the results of BPM-2$^{nd}$ in the same plot. A clear improvement in the results is observed from BPM-2$^{nd}$ to BPM-3$^{rd}$ to BPM-4$^{th}$ with  the value of the slope of the linear fit for BPM-4$^{th}$ (0.986) order data  improved compared to BPM-2$^{nd}$ (0.866) order data. However, the linear-fitted line for BPM-4$^{th}$ is far from the exact $\alpha_{BPM}$ = $\alpha_{DFT}$ line (see Fig.~S\ref{SFig3} (a)). We then compare the calculated time-dependent polarizability trajectory of BPM-2$^{nd}$, BPM-3$^{rd}$ and BPM-4$^{th}$ with DFT in Fig.~S\ref{SFig3} (b). Though the BPM-4$^{th}$-order  shows better results compared to    
BPM-2$^{nd}$,  comparison to DFT results shows that the BPM-4$^{th}$-order still does not correct the  several deviations from the DFT polarizability  trend obtained by BPM-2$^{nd}$ (e.g. see $\alpha_{zz}$ at 13 ps). This mismatch of DFT and BPM results is also reflected in the Raman spectra. The Raman spectra  calculated with BP models of different orders are shown  along with the DFT spectrum   in Fig.~S\ref{SFig3} (c). The peak position of the DFT spectrum is well reproduced by the BPM (2$^{nd}$, 3$^{rd}$ and 4$^{th}$). However, the DFT-calculated peak intensities are not captured well by the BPM (2$^{nd}$, 3$^{rd}$ and 4$^{th}$).

We also compare the results of BPM-3$^{rd}$ and BPM-4$^{th}$ with DFT results for  BTO at 1000 K to check whether the higher order of BPM can improve the accuracy of the BPM results. Fig.~S\ref{SFig4} (a) shows the results of BPM-2$^{nd}$, BPM-3$^{rd}$ and BPM-4$^{th}$ plotted  DFT values. Even though the results are improved overall  with the inclusion of higher order of BPM, the improvement in the accuracy from BPM-2$^{nd}$ to BPM-3$^{rd}$ to BPM-4$^{th}$ is not as significant as that obtained for the 300 K results. This can be seen from the values of the calculated slope of the linearly fitted data points of BPM-2$^{nd}$ (0.778) and BPM-4$^{th}$ (0.778) which are the same. The comparison of time-dependent polarizability trajectory (Fig.~S\ref{SFig4} (b)) from different order of BPM and the corresponding Raman spectra (Fig.~S\ref{SFig4} (c)) also show a noticeable disagreement with the DFT results.

\section{Raman Spectra of BaTiO$_{3}$ single crystal}
The Raman spectra of single crystal BTO for all phases using BPM-2$^{nd}$ are shown in Fig.~S\ref{SFig5}.
